\documentclass[a4paper, 11pt]{article}

\usepackage[utf8x]{inputenc}
\usepackage[top=3cm, left=2.5cm, right=2.5cm, bottom=3cm]{geometry}
\usepackage{graphicx}
\usepackage{amsmath}
\usepackage{amsfonts}
\RequirePackage{amsthm,amsmath,amsfonts,amssymb,bbm,mathtools}
\RequirePackage[round,authoryear]{natbib}
\usepackage{authblk}
\usepackage{comment}
\usepackage{url}

\usepackage{longtable}
\usepackage{verbatim}
\usepackage{bm}
\usepackage[export]{adjustbox}

\newcommand{\wt}{\widetilde}

\newcommand{\loyo}{leave-one-year-out }

\begin{document}

\title{Applications of machine learning to predict seasonal precipitation for East Africa}

\author[1]{Michael Scheuerer}
\author[1]{Claudio Heinrich-Mertsching}
\author[2]{Titike K. Bahaga}
\author[2]{Masilin Gudoshava}
\author[1]{Thordis L. Thorarinsdottir}

\affil[1]{Norwegian Computing Centre, Oslo, Norway}
\affil[2]{IGAD Climate Prediction and Applications Centre (ICPAC), Nairobi, Kenya}

\date{}

\maketitle

\begin{abstract}
\noindent
Seasonal climate forecasts are commonly based on model runs from fully coupled forecasting systems that use Earth system models to represent interactions between the atmosphere, ocean, land and other Earth-system components. Recently, machine learning (ML) methods are increasingly being investigated for this task where large-scale climate variability is linked to local or regional temperature or precipitation in a linear or non-linear fashion. This paper investigates the use of interpretable ML methods to predict seasonal precipitation for East Africa in an operational setting. Dimension reduction is performed by decomposing the precipitation fields via empirical orthogonal functions (EOFs), such that only the respective factor loadings need to the predicted. Indices of large-scale climate variability--including the rate of change in individual indices as well as interactions between different indices--are then used as potential features to obtain tercile forecasts from an interpretable ML algorithm. Several research questions regarding the use of data and the effect of model complexity are studied. The results are compared against the ECMWF seasonal forecasting system (SEAS5) for three seasons--MAM, JJAS and OND--over the period 1993-2020. Compared to climatology for the same period, the ECMWF forecasts have negative skill in MAM and JJAS and significant positive skill in OND. The ML approach is on par with climatology in MAM and JJAS and a significantly positive skill in OND, if not quite at the level of the OND ECMWF forecast.    
\end{abstract}

\section{Introduction} \label{sec:intro}  

The Greater Horn of Africa (GHA), a region here comprised of 11 countries in East Africa, is home to nearly 400 million people\footnote{Population data from \url{https://data.worldbank.org/} (Data for 2023; accessed on August 14, 2024). The 11 countries are Burundi, Djibouti, Ethiopia, Eritrea, Kenya, Rwanda, Somalia, South-Sudan, Sudan, Tanzania and Uganda.} with roughly 75\% of the labor force in the region involved in a smallholder, rain-fed agriculture \cite[]{salami2010} and hydroelectric power accounting for two thirds of power generated \cite[]{IEA2022}. Intraseasonal to decadal rainfall variability thus has a large impact on livelihoods and the region is highly vulnerable to recurrent extreme climate events, with the risks becoming increasingly complex as these hazards are compounded by local and remote political and economic instability \citep{ipcc2022}. Forecasting GHA seasonal rainfall during its different phases is essential for mitigating these impacts. 

The climate of the GHA region is influenced by both complex topographic variations and land-sea gradients \citep[e.g.,][]{anyah2006simulated,sepulchre2006tectonic,Hession2011,thiery2015impact}. Throughout January--February (JF), the northeast monsoon brings dry continental air into the region and rainfall is low except over the southern parts \cite[]{YangEA2015}. Much of the northern and northwestern parts of the GHA have boreal summer monsoon regimes with maximum rainfall during the period of June--September (JJAS). This season accounts for some 50 to 80\% of the rainfall over northern and northwestern agricultural regions \cite[]{Korecha+Barnston2007}. The equatorial part of GHA has two rainy seasons; the bimodal patterns are experienced over many parts of GHA during the "Long Rains" of March--May (MAM) and the "Short Rains" of October--December (OND). 

\begin{figure}[!hbpt]
\centering
\includegraphics[width = 0.45\textwidth]{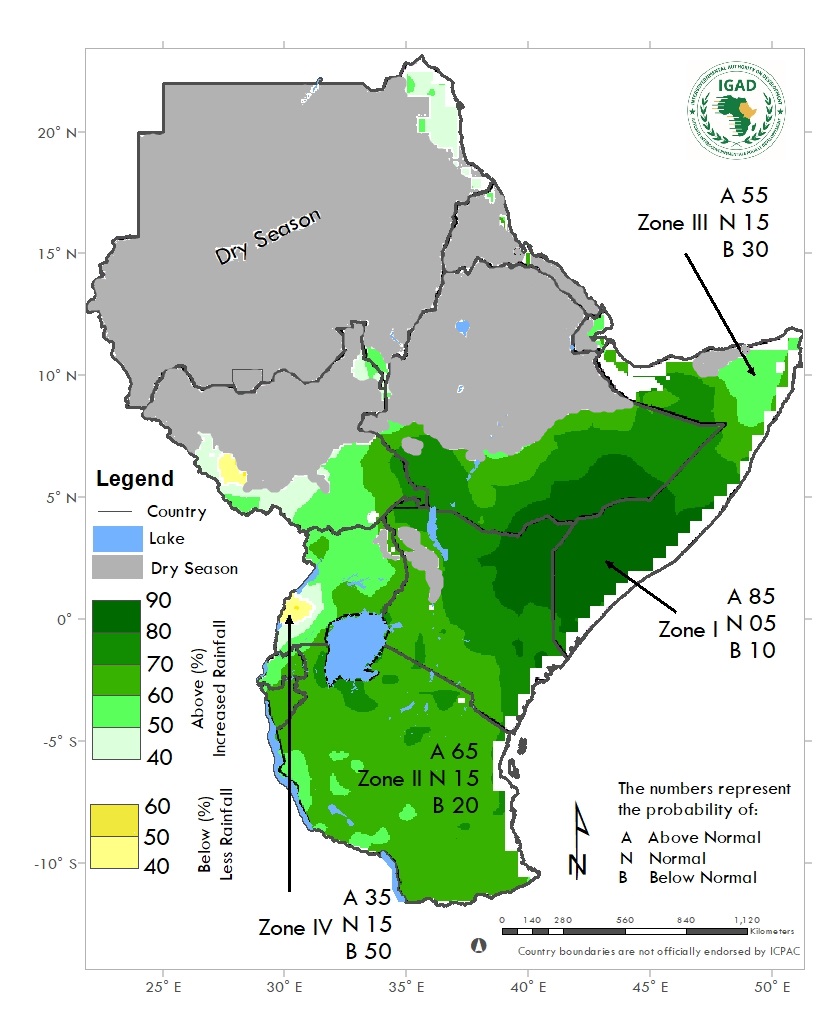}
\caption{Example of tercile prediction for seasonal precipitation. 
The figure shows the forecast for October to December 2023, issued in August 2023 by ICPAC. For most of the region above-normal rainfall is expected with varying degree of confidence.}
\label{fig:tercile_example}
\end{figure}

Climate services over the GHA region are provided by the IGAD Climate Predictions and Applications Centre (ICPAC), a World Meteorological Organization designated regional climate center. Various sub-seasonal and seasonal forecasting products are produced and disseminated on a rolling basis\footnote{\url{https://www.icpac.net/}}. Additionally, seasonal climate information for the three main rainfall seasons is co-produced by climate experts from ICPAC, forecasters from national meteorological and hydrological services (NMHSs), and climate information users, and co-delivered through the Greater Horn of Africa Climate Outlook Forum (GHACOF), a regional user interface platform \citep{walker2019}. GHACOFs are held three times per year, typically in mid-February, the second half of May, and late August, and seasonal forecast products for the MAM, JJAS, and OND seasons are issued, respectively. In this study, we focus on seasonal precipitation forecasts in the operational setting of the GHACOFs. The seasonal forecasts are issued as spatial predictions of tercile probabilities, for above-normal, normal, and below-normal precipitation. The categories are defined by local climatology-terciles, above-normal rainfall corresponding to wetter conditions than observed in 2/3 of all years during the climatology period, for the considered gridpoint and season. An example of such a prediction is shown in Figure \ref{fig:tercile_example}\footnote{\url{https://www.icpac.net/media/images/GHA-Rainfall_OND.original.png}. Accessed on July 1, 2024.}. 

The GHACOFs have been held since 1998 and the early consensus forecasts have been replaced by post-processed dynamical multi-model ensemble forecasts \citep{walker2019, gudoshava2024advances}. The dynamical models are generally found to have skill for the OND season, while predicting seasonal rainfall in MAM and JJAS is more elusive \citep[e.g.][]{BahagaEA2016, Nicholson2017, walker2019, young2020skill}. Alternatively, statistical models identify remote atmospheric and oceanic drivers with teleconnections to the region of interest, to predict conditions in the coming months. The current statistical methods used in operational seasonal forecasting at ICPAC include logistic and linear regression in addition to canoncial correlation analysis. See also \cite{camberlin2002east}, \cite{diro2008seasonal, diro2011teleconnections}, \cite{funk2014predicting}, \cite{Nicholson2014}, \cite{chen2015seasonal}, \cite{colman2020direct} and \cite{kolstad2022lagged} for studies that propose models of this type for the GHA region or a subset thereof. 

The goal of the current study is to develop a seasonal prediction system based on lagged teleconnection observations appropriate for operational use during the GHACOFs. Our focus on developing a system that directly produces probabilistic forecasts in the format of the example in Figure~\ref{fig:tercile_example} has several implications that sets it apart from previous studies. Firstly, we develop models that directly produce spatially gridded tercile forecasts, whereas previous studies commonly focus on prediction of spatial averages, aggregates or indicies \citep[e.g.][]{funk2014predicting, Nicholson2014, deman2022seasonal}. Secondly, an operational context requires a flexible system that is able to produce seasonal forecasts for any month or season and any lead time, automatically selecting appropriate teleconnections from a large pool of potential predictors. Clearly, the skill of such a forecast system will still vary between seasons and will be higher for seasons with strong teleconnections. In seasons with weak teleconnections, even outperforming climatological forecasts can be quite challenging. However, setting up a system that is able to issue prediction for these seasons as well allows us to effectively explore potential teleconnections and directly assess whether they can be used to obtain skillful forecasts.

The need for flexibility in the chosen predictors depending on the situation at hand leads us to look beyond classical statistical regression models \citep[e.g.][]{camberlin2002east, Nicholson2014, kolstad2022lagged} to more flexible machine learning (ML) approaches. ML approaches are becoming increasingly popular in weather and climate modelling and forecasting, see \cite{de2023machine} for a recent review. \cite{he2021sub} compare several data-driven ML approaches for forecasting temperature 3-4 weeks ahead over the contiguous United States and find that extreme gradient boosting (XGBoost) performs best, closely followed by a carefully tuned deep learning approach and the least absolute shrinkage and selection operator (LASSO) method. In a follow up study, \cite{he2022learning} find that while the ML methods are able to match or outperform the skill of dynamical models, the forecasted magnitudes are more conservative. 

For precipitation, \cite{oliveira2023precipitation} provides a recent review of precipitation forecasting aided by ML algorithms, noting the large bulk of published literature on this topic starting with \cite{french1992rainfall}. Studies focusing on the seasonal timescale include \cite{zhou2021comparative},  \cite{khastagir2022efficacy}, \cite{monego2022south}, \cite{perez2022improving} and \cite{yang2022multi}. Here, a variety of ML approaches have been proposed. For example, \cite{lee2024spring} find that the LASSO method performs well in predicting accumulated spring precipitation in South Korea, while \cite{baig2024accurate} show that XGBoost performs well in predicting local monthly precipitation accumulations in the United Arab Emirates. For the Horn of Africa, \cite{deman2022seasonal} aim to predict regional precipitation accumulation during the MAM season. They find that the ML approaches perform on par with the dynamical seasonal forecasting system version 5 (SEAS5) from the European Center for Medium Range Forecasting (ECMWF) only if the test data is incorrectly included in the data set used for selecting predictors. This result nicely demonstrates the importance of careful implementation of ML algorithms with respect to train-test contamination. 

\begin{figure}[!hbpt]
\centering
\includegraphics[width = 0.7\textwidth]{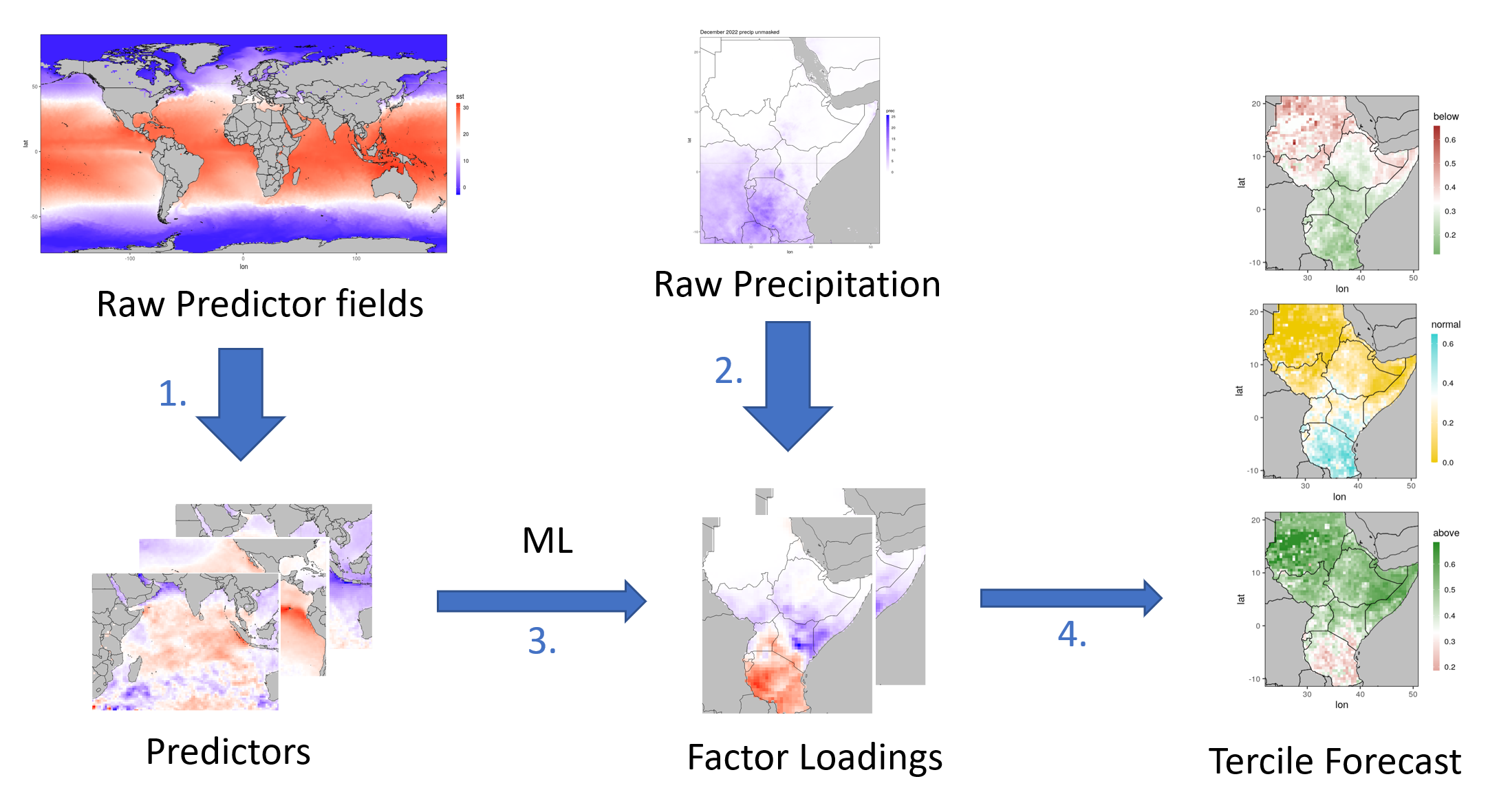}
\caption{Sketch of the different modules of our ML forecasting system. Steps 1 and 2 are data pre-processing and dimension reduction. Step 3 is the central ML method. Step 4 processes the issued low-dimensional predictions into the typically issued tercile forecasts.\label{fig:pipeline}}
\end{figure}

Since we aim to build interpretable ML prediction models, we focus on regression models with regularization techniques, specifically LASSO and elastic nets \citep{ZouHastie2005}, where potential predictors are indices of large-scale climate variability, including the rate of change of individual indices as well as pairwise interactions between different indices. For dimension reduction, we perform empirical orthogonal function (EOF) analysis of the observed precipitation fields and construct a prediction model for the factor loadings, similar to e.g. \cite{camberlin2002east}, \cite{funk2014predicting}, \cite{colman2020direct} and \cite{kolstad2022lagged}. Here, we obtain probabilistic factor loadings predictions that are then further processed to construct spatial tercile forecasts.  An overview of the workflow is given in Figure \ref{fig:pipeline}. 

The implementation and application of the workflow in Figure~\ref{fig:pipeline} raises several questions regarding model choices. We have thus formulated the following research questions: 

\begin{enumerate}
    \item[(i)] Which data product to use as `ground truth' for precipitation? What are the trade-offs between accuracy and availability of data?
    \item[(ii)] How many years of training data to use? While ML methods typically benefit from larger training samples, more training years may require compromises regarding the employed datasets (see (i)) and/or be in conflict with known changes in the relationships between East African rainfall and important teleconnections \citep[e.g.][]{Nicholson2017}.
    \item[(iii)] What level of model complexity is useful/needed? The relationships between different large scale interannual climate drivers and East African rainfall are not independent, and interactions may exist that a linear model would not be able to capture. More complex models, however, are prone to overfitting, given the limited training sample sizes. What are the trade-offs here, and can more advanced regularization techniques make a difference?
    \item[(iv)] Is it better to rely on known teleconnections for which we have a sound meteorological understanding, or does identifying predictors through data-driven approaches allow one to tap into new, unexplored sources of predictability?
    \item[(v)] At which spatial resolution should forecasts be produced? While forecast users often prefer highly localized predictions, forecast skill typically improves with spatial aggregation. To what degree is this true for seasonal predictions of below/above normal rainfall?
\end{enumerate}
We test our ML model with various configurations, which will allow us to discuss these choices. While some of the answers may be specific to the particular forecast task and region studied here, we believe that our study permits useful insights into these questions that generalize well beyond the present setup.

The remainder of the paper is organized as follows. The data sets and the data pre-processing steps are described in the next Section~\ref{sec:data}. In Section~\ref{sec:methods}, we describe the ML algorithms used here, the transformation algorithm for obtaining tercile forecasts from predicted factor loadings, and the forecast evaluation metrics. In Section~\ref{sec:results}, we present the results for the three GHA rainy seasons: MAM, JJAS, and OND. To replicate the GHACOF setting for OND, we consider forecasts issued in August and, therefore, based on information up until the end of July. For MAM, we consider forecasts based on information through January, and the JJAS forecast utilizes information that is available at the end of April. As a reference forecast, we consider ECMWF's SEAS5 forecasts \citep{SEAS5}, which are generally considered best amongst the global prediction systems in this region \citep[e.g.][]{Endris&2021}. Finally, a discussion is given in Section~\ref{sec:disc} and conclusions in Section~\ref{sec:conclusions}. 

\section{Data}\label{sec:data}

This section describes the data sets used in this study as well as all pre-processing steps performed prior to the estimation of the ML prediction model. 

\subsection{Data sets}

\subsubsection{Precipitation data}

We consider two different options for observational data products to train and verify the models against. The first data product used as a proxy for local precipitation amounts is the Climate Hazards group InfraRed Precipitation with a Station (CHIRPS) dataset version 2.0 \citep{FunkEA2015}, downloaded at monthly temporal resolution and 0.05$^\circ \times$ 0.05$^\circ$ spatial resolution and upscaled to 0.5$^\circ \times$ 0.5$^\circ$ resolution. This precipitation product combines in-situ station observations and satellite precipitation estimates, and is heavily used in ICPAC's operations as it is considered to represent rainfall amounts over GHA better than most other data products \citep[e.g.][]{Gebrechorkos&2018,Dinku&2018,Ahmed&2024}.

The CHIRPS data is only available from 1981 onwards, which makes it the limiting factor in our setup regarding training sample size. Therefore, we also consider Global Precipitation Climatology Centre (GPCC) precipitation data \cite[]{schneider&2022}, a gridded rainfall product constructed with data from gauge networks over the globe, which was downloaded at 0.5\textdegree~horizontal resolution for the years 1950-2020. 

\begin{figure}[!hbpt]
\centering
\includegraphics[width = 0.95\textwidth]{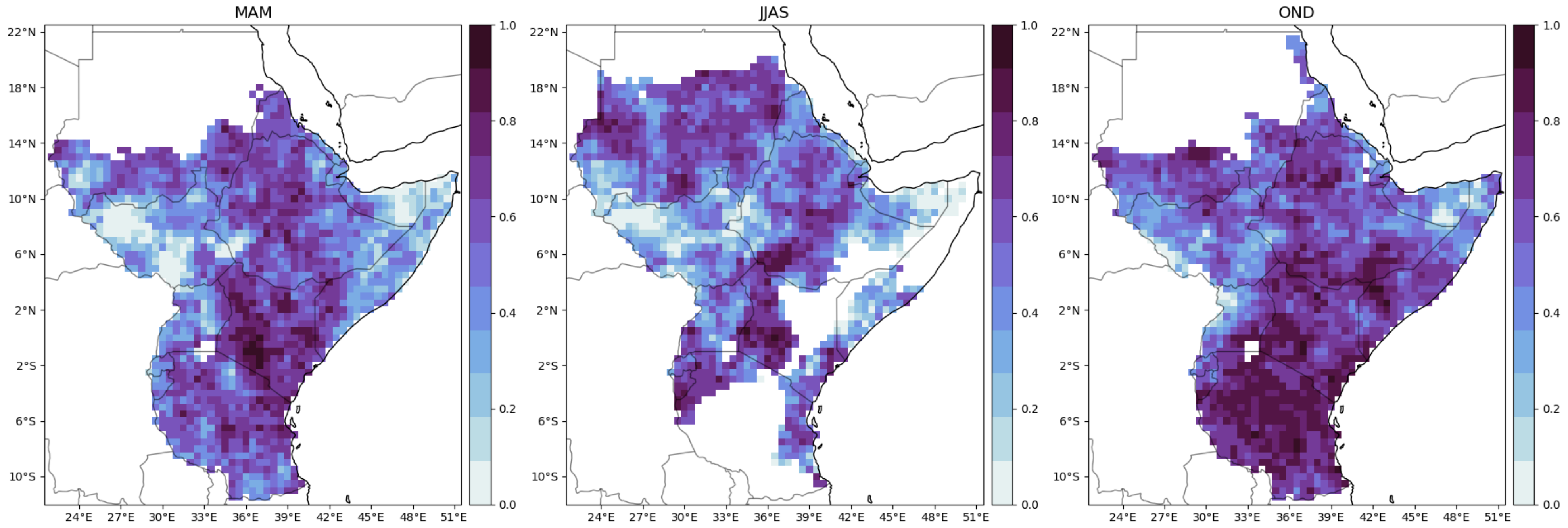}
\caption{Correlations between tercile categories over the time period 1981-2020 calculated from the CHIRPS and GPCC precipitation products for the three seasons considered in this study. Grid points that were too dry to unambiguously determine the tercile category are masked.\label{fig:chirps_vs_gpcc}}
\end{figure}

To investigate the similarity between the CHIRPS and GPCC precipitation products at a relevant scale, we transform each of the data sets over the period from 1981 to 2020 to seasonal tercile categories encoded as $\{-1,0,1\}$ for below normal, normal, and above normal, respectively. For example, if the total accumulated precipitation in MAM in 1981 in a given grid cell is amongst the lowest $1/3$ of the MAM values in that grid cell over the period 1981-2020, it receives a values of $-1$. Figure~\ref{fig:chirps_vs_gpcc} depicts the correlations between the resulting tercile categories, for the three seasons considered in our analysis. In some seasons and regions, the correlations are quite low, which points to a significant disagreement between those two products and emphasizes again the importance of research question 1 formulated above.

For several gridpoints in Somalia and South Sudan, the GPCC data identically repeats a seasonal pattern from 1990 onwards. This is a clear sign of imputed data, presumably due to a lack of available measurements. This affects roughly 5\% of the gridpoints in the considered regions and renders evaluation results for these specific gridpoints questionable, when GPCC data is used. We therefore mostly focus on the use of CHIRPS, both for training and evaluation, and only use GPCC data for analyzing the effect of longer training data.

\subsubsection{Reanalysis data}

Our ML model for forecasting rainfall over GHA relies on predictors derived from global sea surface temperature (SST) and zonal wind data at 200 and 850 hPa levels. For these, we use the ERA5 dataset \cite[]{hersbach&2022}, which was downloaded from the Copernicus Climate Change Service (C3S) Climate Data Store (CDS)\footnote{\url{https://cds.climate.copernicus.eu/}. Accessed on July 1, 2024.} at 0.5\textdegree~horizontal resolution for the years 1950-2020. Previous studies that investigate the skill of seasonal precipitation forecasts derived from statistical-empirical relationships estimated with ML approaches use either observation-based data products \citep{khastagir2022efficacy, zhou2021comparative}, reanalysis products \citep{lee2024spring, monego2022south} or both \citep{deman2022seasonal}. 

\subsubsection{Seasonal forecasts}

As a benchmark for our ML forecasting system we use long-range ensemble precipitation forecasts/hindcasts from ECMWF's SEAS5 system \cite[]{SEAS5}. Monthly aggregates of these data were downloaded from the C3S Climate Data Store at 0.5\textdegree~horizontal resolution for the years 1993-2020. For the hindcast period 1993-2016, 25 ensemble members are available, and for consistency we therefore only use the first 25 (out of 51) ensemble members during the remaining years. As the SEAS5 ensemble is exchangeable, this has the same effect as randomly selecting 25 members from the ensemble.

\subsection{Pre-processing and dimension reduction}\label{subsec:preprocessing}

In this section we describe the processing steps applied to the predictor fields and the precipitation data, which corresponds to steps 1 and 2 of the pipeline shown in Figure \ref{fig:pipeline}. All of these processing steps are applied twice, once using the full data and once taking a leave-one-year-out approach. That is, for the \loyo analysis all data processing such as anomaly calculation, EOF analysis, derivation of indicies and predictor selection is performed separately for each training data fold. The \loyo dataset allows us to fairly assess the skill of our prediction system, see e.g. the discussion in \cite{deman2022seasonal} on the effect of including the test data when selecting correlated predictors. On the other hand, in the \loyo modus, the predictor-selection by our system may vary from year to year, and even the spatial EOF-patterns differ depending on which year is left out. This hinders interpretability of the prediction system. For this reason we also process data and train our model without leaving years out. The results is a single set of selected predictors which can be used for model diagnostics and interpretation.

Seaonal (or monthly, for the predictors) atmospheric and surface climate variables are used to calculate seasonal (monthly) anomalies by subtracting the long-term monthly climatology. However, climate variables are not standardized for each gridpoint (dividing by their standard deviation), since this tends to overemphasize gridpoints with low variability, such as sea surface temperatures close to the freezing point or precipitation close to zero. While detrending can potentially assist the model through separation of long-term or climate change trends and predictor-predictand relationships, the results may be sensitive to the detrending method as trends may, e.g., not be linear in time \citep[e.g.][]{wang2017robust}. Here, the climate variables are not detrended, following the example of, e.g., \cite{Cohen&2019}, \cite{deman2022seasonal}, \cite{lee2024spring}, \cite{monego2022south} and \cite{perez2022improving}. To allow the model to represent trends in precipitation not directly included in any of the potential climate variable predictors, we include the year as a potential predictor.

Many of the processing steps require the selection of a climatological reference period, i.e. a set of years for calculating climatology. For this purpose we generally consider the reference period 1993-2020 (potentially leaving out a single year for generating the \loyo dataset). For these years we have observations from both CHIRPS and GPCC available, as well as fore- or hindcasts from SEAS5. Importantly, the reference period is also used in the evaluation to determine observed terciles. That is, for a fixed gridpoint and season, the observed precipitation is `above normal' if it is higher than 2/3 of the observations between 1993 and 2020, for the same gridpoint and season, and similar for `normal' and `below normal'.  

\subsubsection{Predictors}

We employed 14 predictors derived from ERA5 and SST data to represent anomalous coupled ocean-atmosphere phenomena in the tropical oceans. \cite{BahagaEA2019} revisited the non-stationarity of remote climate anomaly indices and their impact on regional circulation and climate conditions, leading to enhanced or reduced seasonal rainfall over the GHA region. Indices typically constitute (standardized) averages of a climate variable over pre-defined spatial regions (such as the Nino 3.4-region) or differences of such averages (e.g. the dipole mode index (DMI) used for the Indian Ocean dipole). Specifically, we consider sea surface temperature (SST) indices related to El Ni\~no \citep{trenberth1997definition, funk2014predicting}, the Indian Ocean dipole \citep{saji1999dipole} and various oceanic regions as suggested by \cite{Funk&2018}, and winds over the equatorial Indian Ocean at 200 and 850mb \citep{hastenrath2011circulation}. A list of all indices we consider and their definitions are given in the appendix, see Table \ref{tab:indices}. In addition to the indices shown in this table we consider two temporal index differences: n34\_d1 and dmi\_d1. For a given month, n34\_d1 is the difference between the Ni\~{n}o 3.4 index of the current and the previous month, and similar for DMI. The rationale behind this is that the trend (increase or decrease) in these indices could be equally or more important than their value itself \citep{kolstad2022lagged}. We restrict ourselves to lagged versions of these two indices, as they correspond to the two strongest known teleconnections for the GHA region \citep{Nicholson2017, Palmer&2023}.

Standardization of indices is always relative to the climatological reference period 1993-2020.
For deriving indices in \loyo modus, all information of the left-out-year is suppressed in the derivation. In particular, the left-out-year is not used in calculation of the climatological mean (for deriving the anomaly) or standard deviation (used for standardization for several indices). 

\subsubsection{Transformation of precipitation}\label{subsubsec:preprocessing precip}

\begin{figure}
\includegraphics[width = \textwidth]{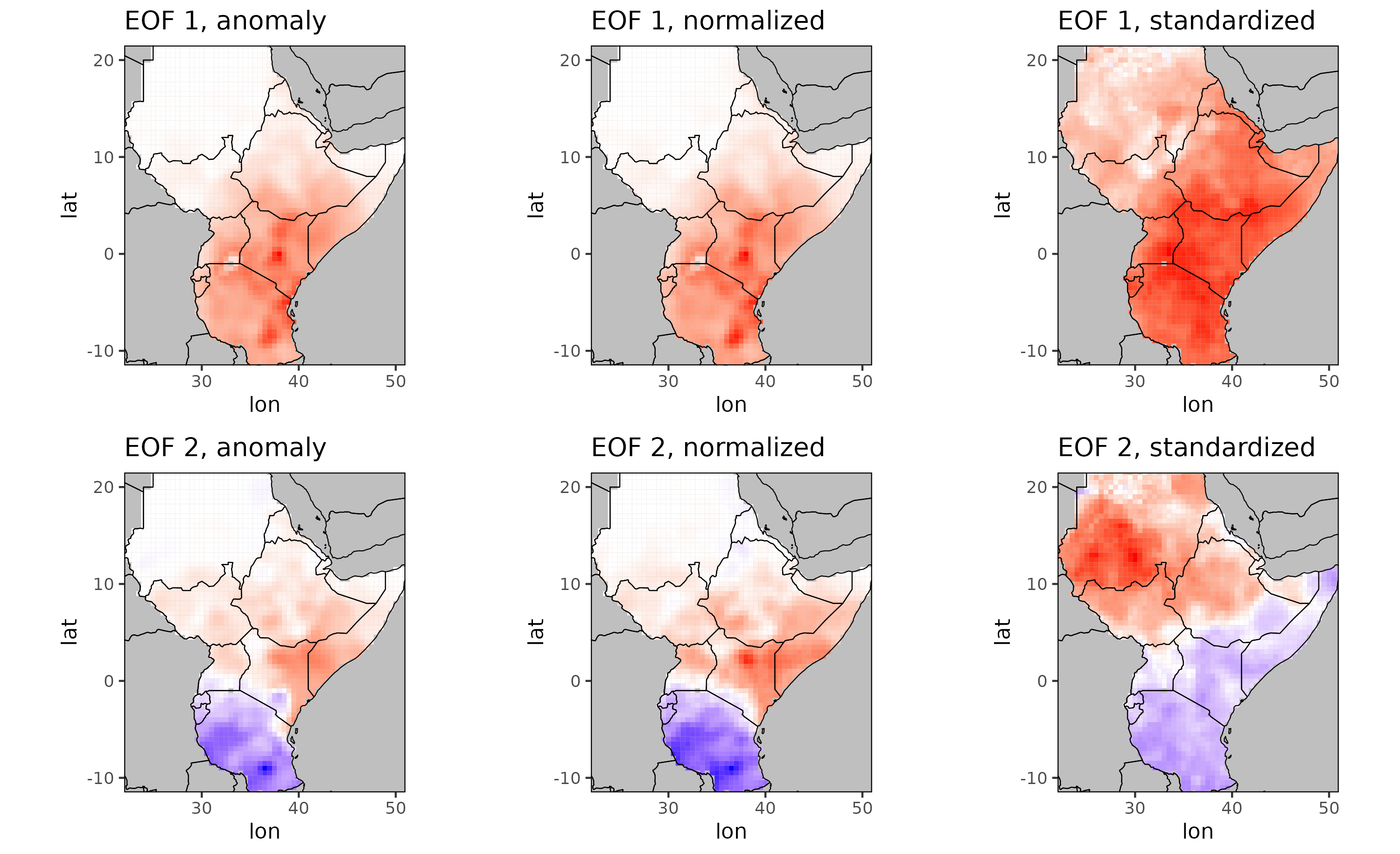}\\
\hspace*{2.5em}
\begin{minipage}{0.3\textwidth}
\includegraphics[width = 0.75\textwidth]{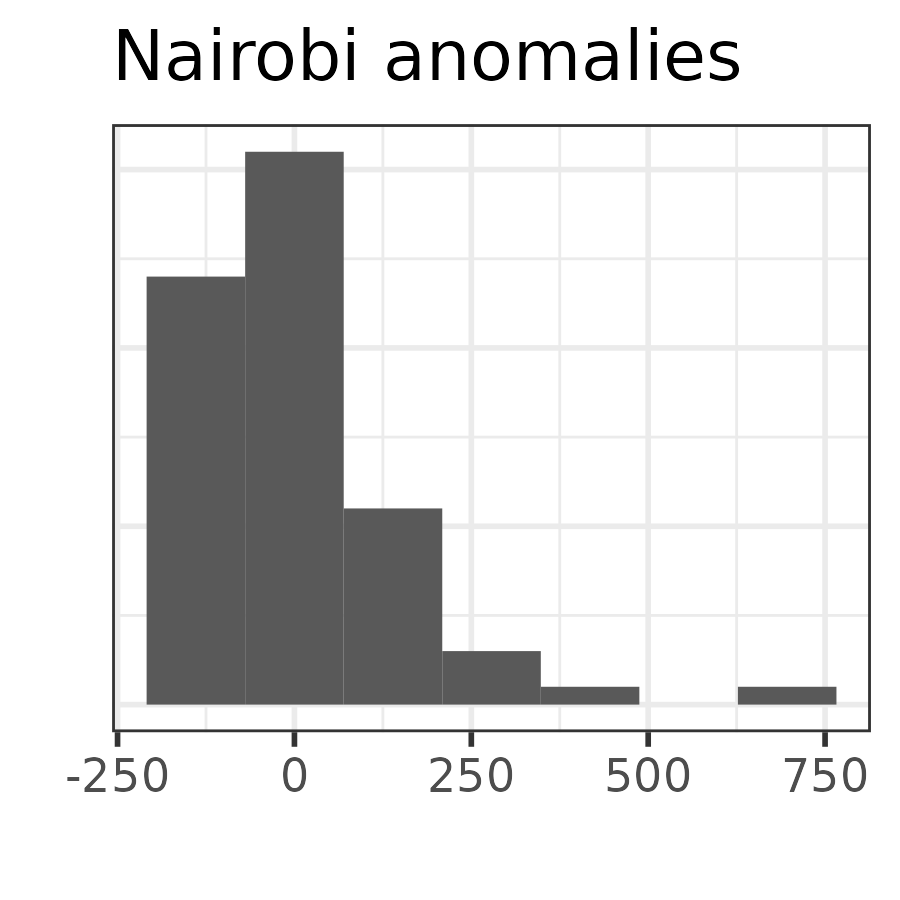}
\end{minipage}
\hspace*{0.8em}
\begin{minipage}{0.3\textwidth}
\includegraphics[width = 0.75\textwidth]{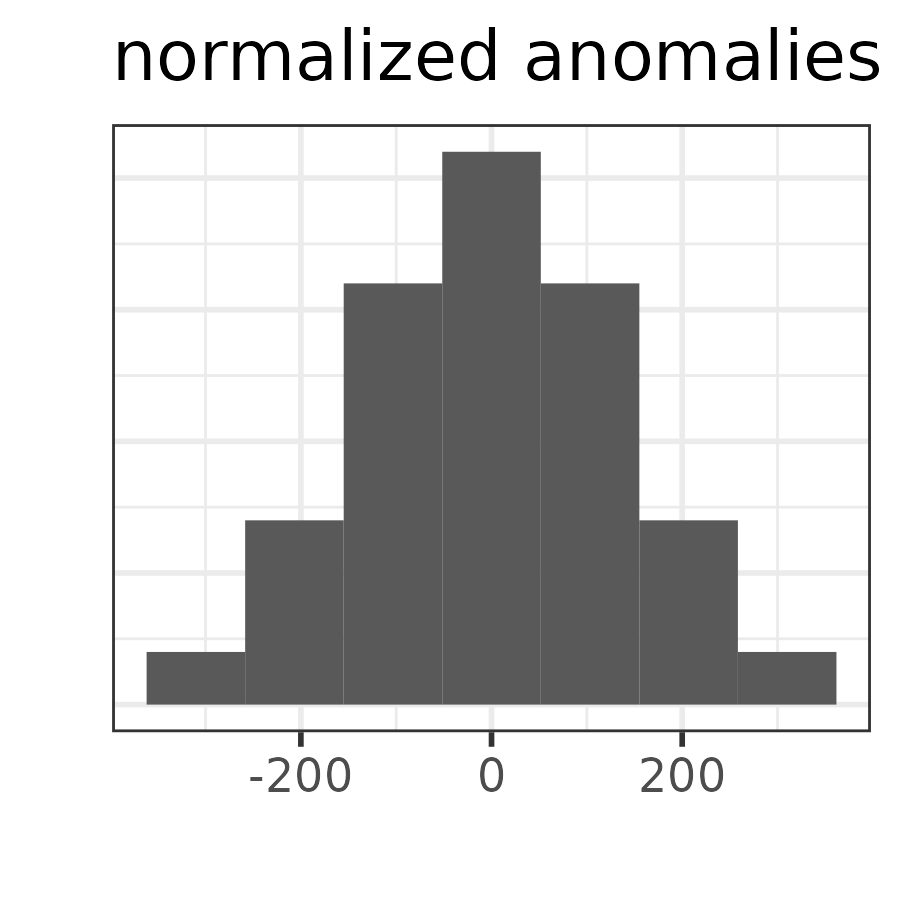}
\end{minipage}
\hspace*{0.8em}
\begin{minipage}{0.225\textwidth}
\includegraphics[width = \textwidth]{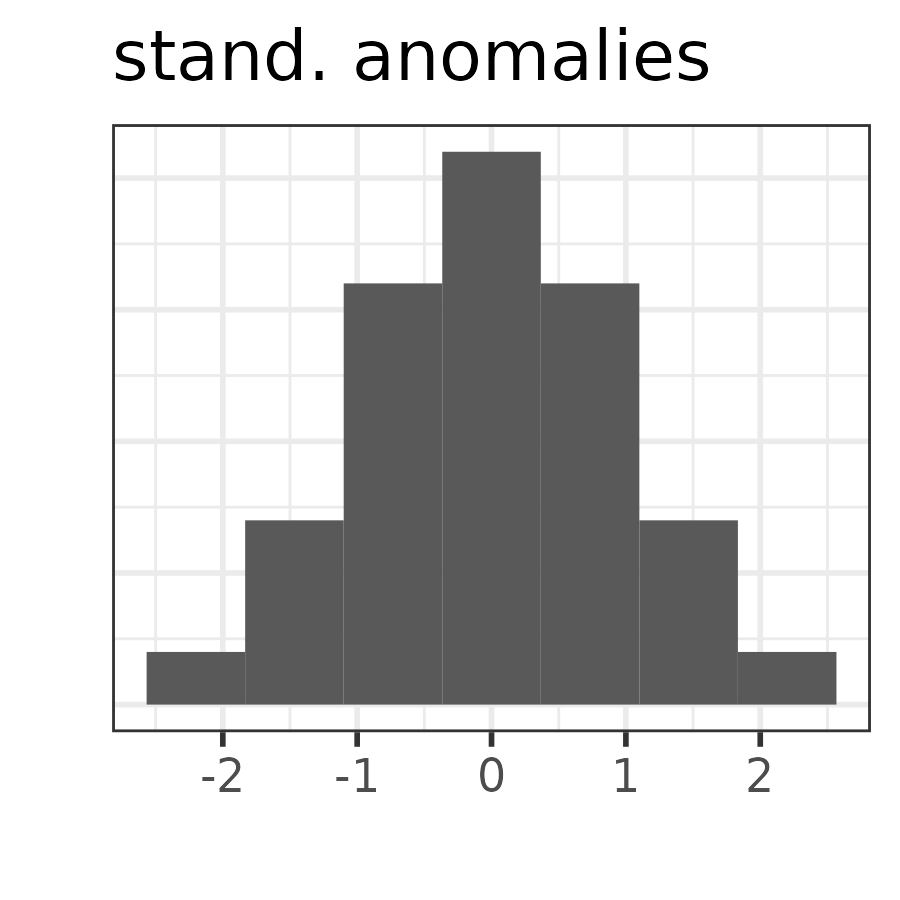}
\end{minipage}
\caption{The top two rows show the first two EOFs for OND-precipitation, from left to right using raw anomalies, anomalies normalized by the transformation given in \eqref{eq:prec_trafo}, and anomalies transformed to a standard normal distribution.
The bottom row shows a histogram of the anomalies for the gridpoint containing Nairobi, before and after transformation. While the transformation in \eqref{eq:prec_trafo} results in similar EOF-patterns as the untransformed anomalies, transformation to standard normal distribution overemphasizes dry regions.\label{fig:prec_eofs}
}
\end{figure}

Precipitation is predicted for all gridpoints contained in the GHA-countries, see Figure \ref{fig:prec_eofs}, by predicting the associated EOF factor loadings. 
Before calculating EOFs, precipitation anomalies are transformed to be normally distributed using quantile mapping. 
Specifically, consider a fixed target season and let $y_{s,t}$ denote the precipitation anomaly for a specific gridpoint $s$ and year $t$.

First, for a year $t$ contained in the reference period 1993-2020, denote by $r_{s,t}$ the rank of $y_{s,t}$ in  $\{y_{s,1993},...,y_{s,2020}\}$. 
We then consider the transformed anomaly
\begin{align}
z_{s,t} := \Phi^{-1}_{\sigma_s}\bigg(\frac{r_{s,t}}{29}\bigg),
\label{eq:prec_trafo}
\end{align}
where $\Phi^{-1}_{\sigma_s}$ is the quantile function of a (centered) normal distribution with standard deviation $\sigma_s$, which is calculated as the empirical standard deviation of $\{y_{s,1993},...,y_{s,2020}\}$. The denominator 29 is the number of years in the reference period plus one.
For a year $t$ not contained in the reference period, the quantiles are interpolated linearly as follows: Let $t_1$ be the year in the reference period with the most precipitation out of all years with less precipitation than $t$, and similarly $t_2$ be the year with the least precipitation of all years with more precipitation than $t$ at location $s$. Then, at location $s$, the year $t$ gets assigned the transformed anomaly
\begin{align}z_{s,t} := (y_{s,t}-y_{s,t_1})\frac{z_{s,t_2} - z_{s,t_1}}{y_{s,t_2} - y_{s,t_1}} + z_{s,t_1},
\label{eq:prec_trafo2}
\end{align}
which is just the value at $y_{s,t}$ of the linear interpolation function that takes values $z_{s,t_1}$ at $y_{s,t_1}$ and $z_{s,t_2}$ at $y_{s,t_2}$.
If the precipitation of year $t$ was lower than for all years of the reference period, we assign the normal quantile $\Phi^{-1}_{\sigma_s}(\frac{1}{30})$, and similarly $\Phi^{-1}_{\sigma_s}(\frac{29}{30})$ if the precipiation was higher than for all reference years.

This transformation allows us to assume that precipitation anomalies are normally distributed at each gridpoint, which we will utilize for the subsequent modelling as described below. Moreover, the transformation keeps the grid point standard deviation unchanged in order to keep spatial consistency.
Using instead a transformation to a standard normal distribution at each gridpoint would put equal emphasis on all gridpoints, regardless of their standard deviation. As a consequence, the prediction system would focus the same amount of attention on very dry gridpoints as on gridpoints that tend to get a lot of precipitation during the considered season, which would be counterproductive. 

The transformation and its impact on EOFs is visualized in Figure \ref{fig:prec_eofs}. As Figure \ref{fig:prec_eofs} shows, there is a slight mismatch between the spatial EOFs of $y$ and $z$. Note that this mismatch does not translate into a spatial error of the prediction system since we only consider the EOFs for $z$. Specifically, we train our prediction model to predict factor loadings of $z$ and then translate them to tercile probabilities using the climatological distribution of $z$. In the context of the forecasting system, it is constructive to losely interpret the EOFs of $z$ as ``patterns of attention'' of the prediction model. That these patterns are close to the EOFs of $y$ indicates that the attention of the model aligns with patterns of maximal variability of precipitation. 

\subsubsection{Empirical orthogonal function (EOF) analysis of transformed precipitation}\label{subsec:EOF analysis}

Let $\mathcal S$ denote the collection of gridpoints for which we issue predictions, cf. Figure \ref{fig:chirps_vs_gpcc}.
EOF analysis constructs orthogonal patterns of maximal variability from a set of vectors, in our case the transformed precipitation fields $\bm{z}_t := (z_{s,t})_{s\in\mathcal{S}}$ over a collection of years $t$ \citep{Jolliffe2016}.
We calculate EOFs generally over the reference period $\mathcal T := \{1993,...,2020\}$, where we additionally leave out the target year for prediction in cross-validation mode.

The EOFs are orthonormal vectors $\bm{\zeta}_1 := (\zeta_{s,1})_{s\in\mathcal{S}}, \bm{\zeta}_2,...$ which are constructed to maximize the variability of $\bm{z}_t$ in the subspace spanned by $\bm{\zeta}_1,...,\bm{\zeta}_k$. 
For a given year $t$, the scalar product $a_{i,t} := \bm{\zeta}'_i \bm{z}_t$ is called the $i$th \emph{factor loading}.
The vector $\bm{a}_t=(a_{1,t},...,a_{k,t})'$ forms a $k$-dimensional representation of the $|\mathcal S|$-dimensional transformed anomaly $\bm{z}_t$ that is in a sense statistically optimal. In particular, the first few factor loadings typically represent large-scale variations, whereas higher factor loadings represent more and more local patterns which tend to be less predictable. 
Consequently, our method considers the decomposition of $\bm{z}_t$ into the first few factor loadings  and residual fields $(\epsilon_{s,t})_{s\in\mathcal{S}, t\in\mathcal{T}}$, which we regard as unpredictable "noise":
\begin{equation}
    z_{s,t} = \sum_{i=1}^k a_{i,t} \, \zeta_{s,i} + \epsilon_{s,t}, \quad s\in\mathcal{S}, \; t\in\mathcal{T}. \label{eq:precip-field-decomposition}
\end{equation}
By construction, we can assume that the random vector $(\epsilon_{s,t})_{s\in\mathcal S}$ is distributed as $\mathcal N(0,\wt\Sigma)$ with some spatial covariance matrix $\wt\Sigma$. Only the diagonal entries $\wt\sigma_{s}^2$ of $\wt\Sigma$ are needed for building the prediction model below; these can be estimated as the difference between $\sigma_{s}^2$ and the variance at $s$ explained by the $k$ factor loadings.
%\FloatBarrier

\section{Methods}\label{sec:methods}

This section describes in detail the ML forecasting system, the generation of tercile forecasts and the evaluation of such forecasts. The description of the ML system refers to a fixed forecast season and initialization month, with the respective indices omitted for clarity of notation. 

\subsection{Machine learning algorithm}\label{subsec:ML}

For a fixed year $t$, our model issues a multivariate normal prediction for the joint factor loadings defined in \eqref{eq:precip-field-decomposition}. That is, we assume a conditional distribution for $\bm{a}_t$ of the form
\[\bm{a}_t \sim \mathcal N\big(\bm{\mu^{(a)}}_t,\Sigma^{(a)}\big),\]
where $\bm{\mu^{(a)}}_t$ is the mean prediction from the ML model which depends on the observed predictors in the year $t$. While $\Sigma^{(a)}$ also weakly depends on $t$ due to the \loyo setup, we ignore this dependence in the model description. In principle, any supervised learning algorithm can be trained to predict the mean $\bm{\mu^{(a)}}_t$ and the covariance matrix $\Sigma^{(a)}$ of the prediction error. Below, we describe the details of the particular algorithm we have chosen, where $\bm{\mu^{(a)}}_t$ and $\Sigma^{(a)}$ are estimated separately. 

\subsubsection{Predictor pre-selection}\label{subsec:predictor-pre-selection}

As a first step to predicting $\bm{a}_t$ for a given year $t$, we pre-screen the set of candidate predictors constructed in Section~\ref{subsec:preprocessing} and remove those that have no correlation with any of the components of $\bm{a}_t$. Specifically, we consider $p$-values of a two-sided test for the empirical correlations $\rho_1,\ldots,\rho_k$ between each of the components of $\bm{a}_t$ and a given predictor being significantly different from zero. In the test, we require a higher significance standard for higher EOFs and use significance levels
\begin{equation}
    \alpha_i = 0.1 \cdot \sqrt{\lambda_i / \sum_{j=1}^k \lambda_k}, \qquad i = 1, \ldots,k.
\end{equation}
where $\lambda_i$ is the variance of the $i$th factor loading. Predictors for which the null hypothesis $\rho_i=0$ cannot be rejected for at least one $i \in \{1,\ldots,k\}$ will not be considered further.

\subsubsection{Features used as input to the ML system}\label{subsec:feature-construction}

So far, we have used the term "predictors" for the meteorological variables that can potentially provide information about future precipitation amounts over GHA. In the ML literature, the term "features" is commonly used for the set of inputs to a machine learning system. Here, we construct such features from the original, pre-selected predictors, by considering predictor interactions in addition to the predictors themselves. 

The interactions we consider involve dummy variables with values in $\{0,1\}$ of the form $\bm{1}\{\text{predictor } u \text{ below normal}\}$ and $\bm{1}\{\text{predictor } u \text{ above normal}\}$, where $\bm{1}$ denotes the indicator function. Thereby, we hope to capture situations where e.g.\ one meteorological variable has a notable influence on the prediction target only when another one is in a certain "active" state. Interactions of the form $v \cdot\bm{1}\{$predictor $u$ below normal$\}$ and $v\cdot\bm{1}\{$predictor $u$ above normal$\}$ are considered both across different meteorological predictors $u, v$ as well as the same predictor (i.e.\ $u=v$). In the latter case, they represent a non-linear transformation similar to the rectified linear unit \citetext{ReLU, \citealp{Fukushima1969}} activation function commonly used in neural networks \cite[]{Glorot&2011}, while preserving model interpretability. 

For any year $t$, we compose a feature vector $\bm{x}_t$ consisting of the original (pre-selected) predictors, their interactions, and a further feature defined as $(t-2000)/10$ in order to capture possible trends over time. After stacking the transposed vectors $\bm{x}_t^\prime$ and  $\bm{a}_t^\prime$ for all training years $t\in\mathcal{T}_{tr}$ row-wise into a design matrix $\bm{X}$ and a response matrix $\bm{A}$, we can formulate a  multi-task linear regression model
\begin{equation}
  \bm{A} = \bm{X}\bm{B} + \tau \label{eq:linear-regression-model},
\end{equation}
in which $\bm{B}$ is the matrix of regression coefficient (with one column for each of the $k$ response variables) and $\tau$ is the error matrix corresponding to the variability in $\bm{A}$ not explained by the feature values in $\bm{X}$.

\subsubsection{Regularization}\label{subsec:regularization}

Despite the pre-selection of predictors, there is potentially a very large number of features, especially when interaction terms are considered which square (and double) the number of features compared to a basic model where only the original predictors are considered. With our limited training sample size, a standard least squares estimation of $\bm{B}$ in \eqref{eq:linear-regression-model} would almost certainly lead to severe overfitting and poor out-of-sample performance, i.e.\ poor generalization of the fitted model to years not in $\mathcal{T}_{tr}$.

To deal with this situation, i.e., many features and limited training sample size, regularization techniques like ridge regression \cite[]{HoerlKennard1970a,HoerlKennard1970b} or LASSO \cite[]{Tibshirani1996} have been proposed to shrink the regression coefficients towards zero with the goal of reducing the variance (i.e., uncertainty) of the estimated regression coefficients. While the former was more targeted at the (overfitting) challenges that come with highly correlated features, the latter focuses on the problem of selecting the most important features among a possibly large set of candidates. \cite{ZouHastie2005} later introduced the concept of "elastic net" regularization which includes the penalty terms from both ridge regression and LASSO and was demonstrated to combine their respective advantages by performing feature selection while avoiding LASSO's tendency to select only one single feature out of a group of highly correlated features.

Formally, in elastic net regression, an estimate $\hat{\bm{B}}$ of the parameter matrix $\bm{B}$ in our multi-task regression setting \eqref{eq:linear-regression-model} is found by minimizing the objective function
\begin{equation}
  g_{\lambda_1,\lambda_2}(\bm{B}) = \|\bm{A}-\bm{X}\bm{B}\|_F^2 + \lambda_2\|\bm{B}\|_F^2 + \lambda_1\|\bm{B}\|_{21} \label{eq:elastic-net-multi-task-objective},
\end{equation}
with $\|\bm{B}\|_F^2 = \sum_{i=1}^d\sum_{j=1}^k B_{ij}^2$ ("Frobenius norm") and $\|\bm{B}\|_{21} = \sum_{i=1}^d \sqrt{\sum_{j=1}^k B_{ij}^2}$.
Unlike a model fitting strategy which estimates the parameters for each of the $k$ components separately, this multi-task model fitting approach selects the same features for all $k$ components. The regularization hyper-parameters $\lambda_1,\lambda_2$ are selected through a further cross-validation loop that runs inside the leave-one-year-out cross validation described above. It partitions the training data further into 5 different folds, keeping years together in pairs of two while stepping through the full data set in steps of 10 years to make sure that each decade is represented equally in each of the 5 folds. For a given pair $(\lambda_1,\lambda_2)$, the model is fitted to 4 of the 5 folds and validated on the left-out fold; the hyper-parameter pair with the lowest mean squared error, averaged over the 5 validation folds, is then used for re-fitting the model to the full training data set. As a baseline approach, we use multi-task LASSO, which corresponds to the special case where $\lambda_2=0$, while $\lambda_1$ is determined as described above.

\subsubsection{Quantifying forecast uncertainty}\label{subsec:uncertainty}

Given the estimated parameter matrix $\hat{\bm{B}}$ and a feature vector $\bm{x}_t$, a point forecast $\hat{\bm{\mu}}_t^{(a)}$ of $\bm{a}_t$ is obtained as
\begin{equation}
  \hat{\bm{\mu}}_t^{(a)} = \hat{\bm{B}}^\prime \bm{x}_t \label{eq:linear-prediction-model}.
\end{equation}
The covariance matrix $\bm{\Sigma}^{(a)}$ of the prediction error can be estimated from the residuals $\bm{a}_t-\hat{\bm{\mu}}_t^{(a)}$ of the fitted values across all training years $\mathcal{T}_{tr}$ via
\begin{equation}
 \hat{\bm{\Sigma}}^{(a)} = \frac{1}{|\mathcal{T}_{tr}|-{\it df}} \sum_{t\in\mathcal{T}_{tr}} \big(\bm{a}_t-\hat{\bm{\mu}}_t^{(a)}\big)\big(\bm{a}_t-\hat{\bm{\mu}}_t^{(a)}\big)^\prime \label{eq:covariance-estimation}.
\end{equation}
Here, $df$ denote the degrees of freedom of the statistic $\bm{\Sigma}^{(a)}$. Subtracting the degrees of freedom in the denominator is required to obtain an unbiased estimator.
Unlike in standard regression, where ${\it df}$ is simply the number of columns of $\bm{X}$, the elastic net regularization makes it that ${\it df}$ is itself a random quantity that needs to be estimated along with $\hat{\bm{B}}$. 

For the LASSO case where $\lambda_2=0$, \cite{Zou&2007} show that  ${\it df}$ is number of non-zero coefficients in the (single-objective version of) linear model~\eqref{eq:linear-prediction-model}. In an earlier presentation \cite[]{enet_talk}, Zou and Hastie discuss the more general case where $\lambda_2\geq0$ and suggest an estimate for ${\it df}$ given by 
\begin{equation}
  \widehat{\it df} = \mathrm{tr}(\bm{H}_{\lambda_2}(\mathcal{A})), \qquad \bm{H}_{\lambda_2}(\mathcal{A}) = \bm{X}_{\mathcal{A}}(\bm{X}_{\mathcal{A}}^\prime \bm{X}_{\mathcal{A}} + \lambda_2\bm{I})\bm{X}_{\mathcal{A}}^\prime \label{eq:estimated-degrees_of_freedom},
\end{equation}
where $\mathrm{tr}$ denotes the trace operator, $\bm{I}$ denotes the identity matrix, and $\bm{X}_{\mathcal{A}}$ denotes the sub-matrix of $\bm{X}$ corresponding to the subset of features with a non-zero coefficient.

\subsection{Tercile forecasts}\label{subsec:tfcs}

For a given gridpoint and a given season, a tercile forecast takes the form of three probabilities $p_{b},\,p_n,\,p_a$ (summing to 1) for the below-normal, normal, and above-normal category, respectively. The three categories are defined by climatology terciles and, consequently, a climatological prediction always issues the probabilities $p_b = p_n = p_a = 1/3$. This forecast is frequently used as reference, for example to calculate skill scores.

From \eqref{eq:precip-field-decomposition}, we obtain a probabilistic forecast for the transformed precipitation amount $z_{s,t}$ given by 
\begin{align}
z_{s,t} \sim \mathcal{N}\bigg(\sum_{i = 1}^k \hat{\mu}^{(a)}_{i,t}\zeta_{s,i}, \sum_{i,j = 1}^k \zeta_{s,i}\hat{\Sigma}^{(a)}_{ij}\zeta_{s,j} + \wt\sigma_{s}^2\bigg),\label{eq:predictive_dist}
\end{align}
where $\wt\sigma_{s}^2$ is the variance of the residual fields in the EOF analysis described in subsection~\ref{subsec:EOF analysis}. The climatological distribution of $z_{s,t}$ is $\mathcal N(0,\sigma_s^2)$, and thus we can derive a tercile forecast from the probabilistic forecast of $z_{s,t}$ directly by comparing our prediction to the climatological distribution.
To be precise, let $F_{s,t}$ denote the cumulative distribution function (c.d.f.) of the normal distribution in \eqref{eq:predictive_dist}. The predicted probabilities for below- and above normal rainfall are then $F_{s,t}(\Phi^{-1}_{\sigma_s}(1/3))$ and $1 - F_{s,t}(\Phi^{-1}_{\sigma_s}(2/3))$, respectively. In particular, we obtain predictions of tercile probabilities directly from the prediction of the transformed precipitation without needing to invert the transformation in \eqref{eq:prec_trafo2}.

Tercile forecasts can easily be derived from ensemble forecasts, by comparing the ensemble forecast for a given year with the forecast climatology. To be specific, for a given grid point and a given season, tercile boundaries are calculated from all ensemble members for this grid point and period, considering all years used as climatological reference period. For a given year, the predicted probabilities are then the fractions of ensemble members falling into the respective tercile. Using the model climatology to define the tercile boundaries automatically removes possible systematic biases without any need for explicit bias correction. In fact, this way of relating terciles of the ensemble forecast to climatology terciles constitutes a special case of statistical post-processing by quantile mapping \citep[e.g.][]{zhao2017suitable}.

\subsection{Tercile forecast evaluation}\label{subsec:eval}

For evaluating the tercile forecasts, we employ proper scoring rules \citep[e.g.][]{GneitingRaftery2007} which evaluate both reliability (i.e., whether predicted probabilities align with observed frequencies) and resolution (i.e., the forecast's confidence about the outcome) of a forecast. Specifically, we use the (multicategory) Brier Score (MBS), defined as 
\[MBS(y,(p_b,p_n,p_a)) := (p_b-\mathbf 1_b(y))^2 + (p_n - \mathbf 1_n(y))^2 + (p_a - \mathbf 1_a(y))^2,\]
where $\mathbf 1_b(y)$ is the indicator whether the observation $y$ falls into the below category, and similarly for $\mathbf 1_n$ and $\mathbf 1_a$. The MBS is negatively oriented, in the sense that a lower score is associated with a better prediction. In practice, a prediction system is evaluated by calculating the average score over many forecast-observation pairs. The average score is then frequently transformed into a skill score relative to the climatological forecast. A perfect forecast system attains a skill score of 1 and the climatological reference forecast itself gets a skill score of 0. Skill scores above 0 indicate better performance than the climatological forecast, and higher values indicate better forecasts. Note that skill scores should, in general, not be compared across studies that are based on different data sets which might have different climatological distributions, see \citet{GneitingRaftery2007} and references therein. 

Evaluation of tercile forecasts is possible at every gridpoint, including gridpoints that receive almost no precipitation during the season under study. However, the resulting scores are not reliable in grid points too dry to unambiguously determine the tercile category. For the evaluation, we thus apply the spatial masks shown in Figure~\ref{fig:chirps_vs_gpcc}. 

\section{Results}\label{sec:results}

Here, we present the results of our investigations into seasonal precipitation forecast skill in the GHA region for the three rainy seasons MAM, JJAS and OND. Following the schedule of the GHACOFs, the forecasts may use data available by mid-February, mid-May, and mid-August, respectively. For the SEAS5 ensemble, this includes the runs initialized at the beginning of the respective month, while for the ML based forecasts only analyses from the month(s) prior can be used. The evaluation is performed over the period 1993-2020 and skill scores are evaluated relative to the climatological references forecast over the same time period, where the to-be-evaluated year is removed from the climatological distribution.  

\subsection{Forecast skill of ECMWF seasonal precipitation forecasts}\label{subsec:ecmwf}

\begin{figure}[!hbpt]
    \centering
    \includegraphics[width = 0.8\textwidth]{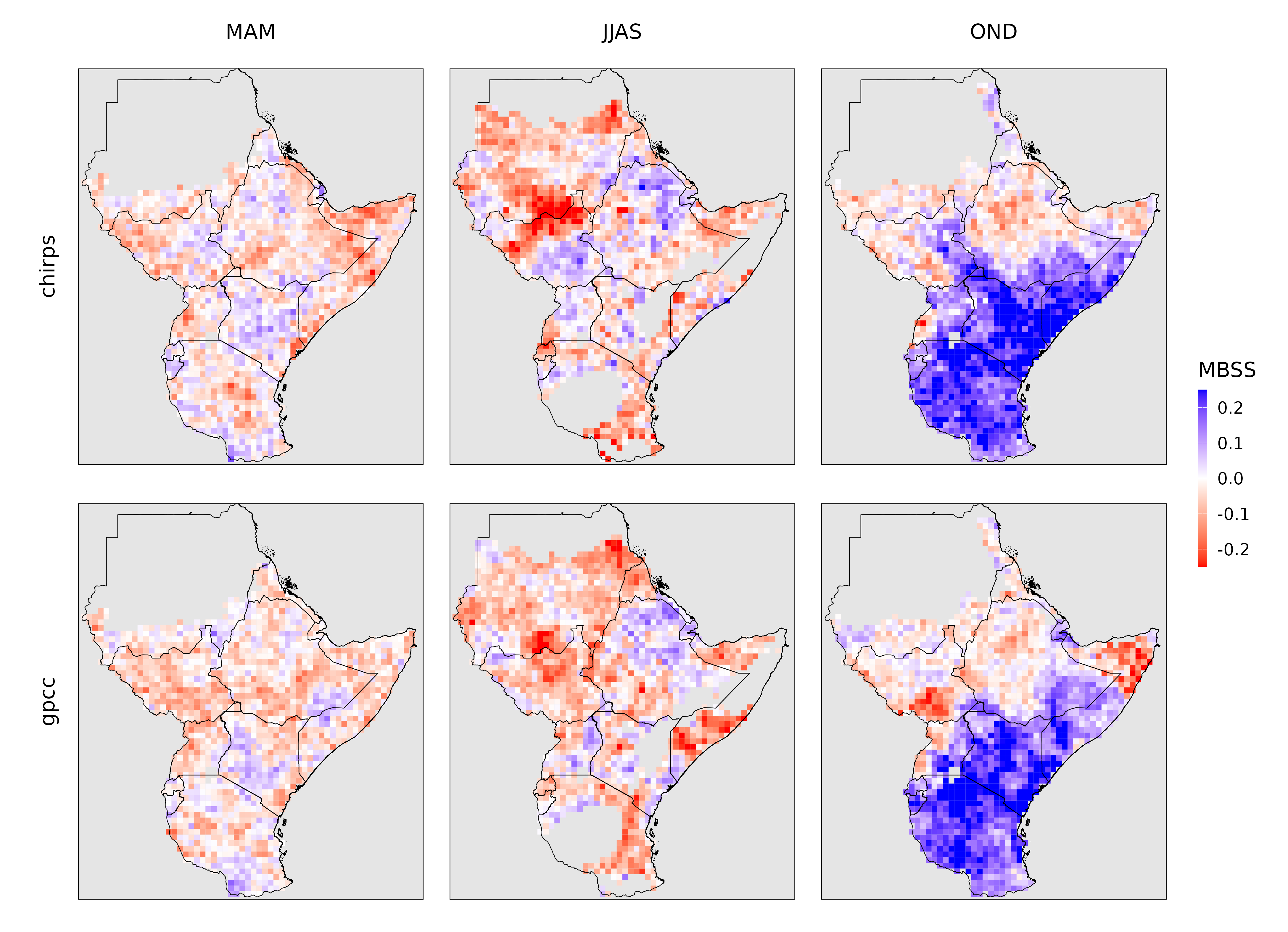}
    \caption{Multicategory Brier skill scores (MBSS) over the period 1993-2020 of SEAS5 tercile forecasts evaluated against CHIPRS (top row) and GPCC (bottom row), for the three considered seasons.}
    \label{fig:mbss-map-ecmwf}
\end{figure}

To establish a benchmark for the forecast skill we can expect over GHA, we first look at MBSS maps for tercile forecasts derived from ECMWF's SEAS5 ensemble (see Figure~\ref{fig:mbss-map-ecmwf}). For MAM and JJAS, the SEAS5 forecast skill is negative at the majority of grid points. This is a result of the limited predictability of GHA precipitation usually found for these seasons and a tendency of the ensemble to be overconfident, which in combination leads to the ensemble representing the range of possible outcomes worse than climatology. For OND, on the contrary, the ECMWF ensemble skill is positive over most of GHA, especially over the regions that receive a substantial amount of their annual precipitation during that season.

The MBSSs are noticeably lower when the forecasts are verified against GPCC data, especially over the regions where substantial data imputation in the construction of the GPCC product had to be performed as discussed in Section~\ref{sec:data}. We see this as evidence for the higher quality of CHIRPS compared to GPCC, and will therefore focus on verification against CHIRPS. Verification (and training) against GPCC will be revisited in Section~\ref{subsec:training-sample-size} though, when we address the research question related to the tradeoff between training sample size and data quality.

\subsection{Forecast skill of the machine learning based seasonal precipitation forecasts}\label{subsec:ml-skill}

We now look at MBSSs of tercile forecasts produced by different variants of the ML approach proposed in Section~\ref{sec:methods}. Specifically, we consider regression models with the indicies listed in Table~\ref{tab:indices} and various degrees of regularization: elastic nets where the ratio of the two penalty terms is fixed (indices\_enfixed) or optimized (indices\_enopt), and LASSO (indices\_lasso). For this baseline model (LASSO), we also consider adding index interaction terms (indices\_lasso\_iat) and a fully data-driven variant where the predictors are factor loadings obtained through EOF analysis of SSTs in different oceanic regions (sst\_fls\_lasso). 

\begin{figure}[!hbpt]
    \centering
    \includegraphics[width = 0.9\textwidth]{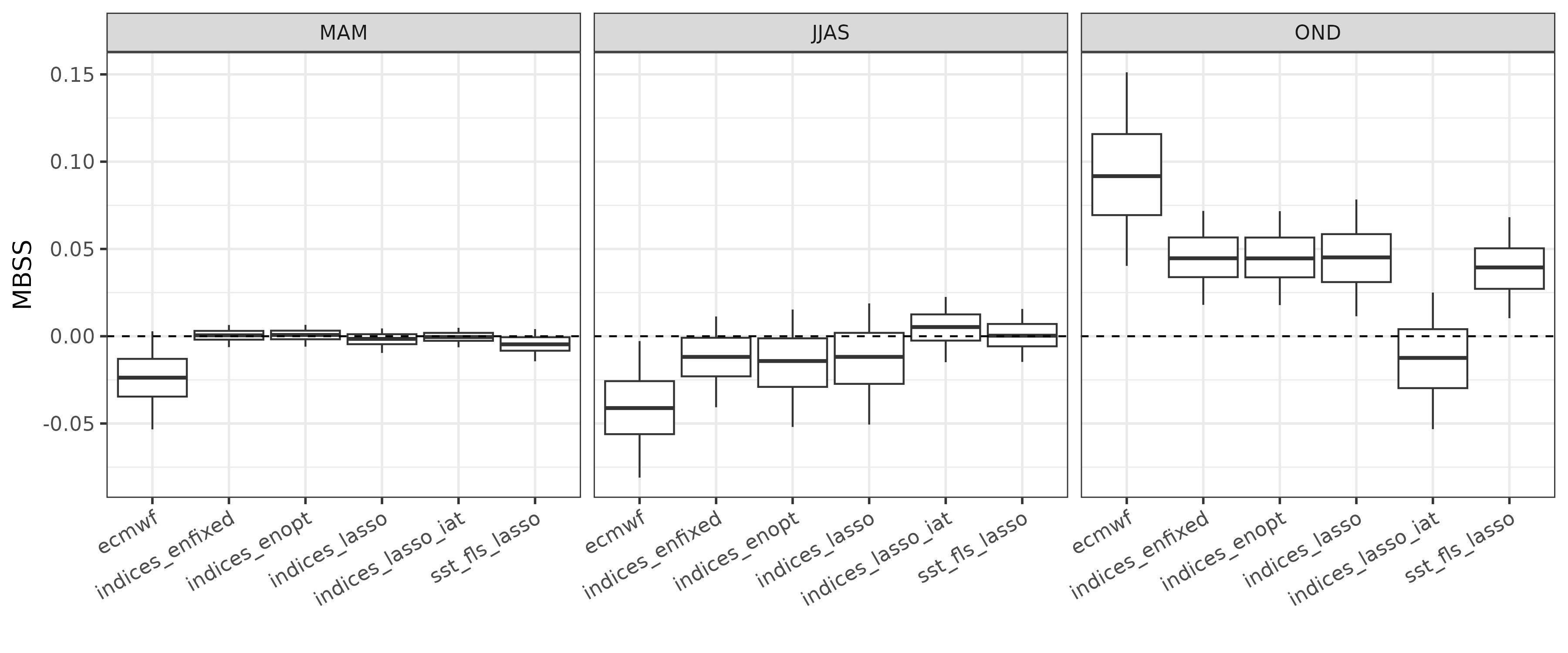}
    \caption{MBSSs for seasonal precipitation forecasts by ECMWF's SEAS5 ensemble and various configurations of our ML forecast system. The boxplots show bootstrap resamples of the mean score, see text for details. The forecasts are verified against CHIRPS over the period 1993-2020. The lead times we consider are determined by the typical GHACOF dates; the forecasts may use information up to one month prior to start of the respective season.}
    \label{fig:skill_bootstrapped_chirps}
\end{figure}

In order to get a quick overview, scores are aggregated over the masked regions shown in Figure~\ref{fig:mbss-map-ecmwf}. To get an idea of the magnitude of sampling variability bootstrapping of the spatially averaged scores obtained for each year was performed, and the range of bootstrapped outcomes is represented as a boxplot in Figure~\ref{fig:skill_bootstrapped_chirps}.
Specifically, we repeatedly sample the 28 verification years (1993-2020) 1000 times with replacement, and calculate the average score over the resampled years. The variability of the resulting scores provides a measure of how robust differences between models are and to what extend they vary from year to year. The whiskers of the boxes were chosen to represent the 5th and 95th percentile. If the zero skill line is outside of that range, this can be interpreted as a rejection of a (one-sided) hypothesis of greater or equal, or lesser or equal skill than climatology at a 5\% significance level.

We start with a discussion of the results obtained for the OND season, which is the most predictable for this region as noted in the literature \citep[e.g.][]{BahagaEA2016, walker2019, young2020skill} and confirmed in Figure~\ref{fig:mbss-map-ecmwf}. Indeed, the MBSS attained by most of the variants of our ML approach studied here is positive, though considerably lower than the MBSS of the ECMWF forecast. The scores are very similar across the model variants that differ only in their utilized regularization approach, in particular there is little evidence to suggest that the elastic net provides an advantage over LASSO, regardless of whether the ratio between the two penalty terms is fixed or optimized. Likewise, the data driven EOF-based predictors perform similar to the predefined indices, which suggests that the data driven approach can identify the relevant sources of predictability but is unable to identify additional sources of predictability that would make a crucial difference. Including predictor interactions as additional features, on the contrary, deteriorates the OND forecast performance of the ML approach. Apparently, the benefit of increased flexibility is outweighed by the need for stronger regularization due to the massive increase of model parameters, which dampens the predictable signal and compromises forecast performance.

\begin{figure}[!hbpt]
    \centering
    \includegraphics[width = 0.9\textwidth]{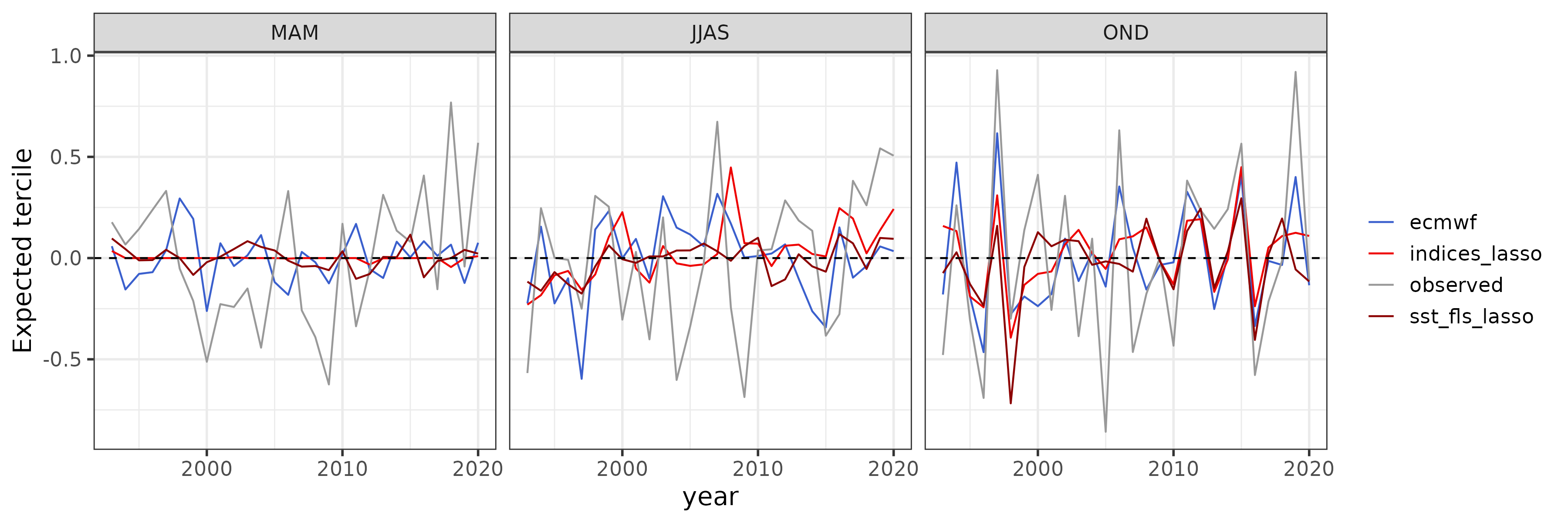}
    \caption{Expected terciles over the GHA region for each season and year, where a positive value indicates an overall wetter than normal conditions with a value of one indicating wetter than normal in every grid cell, and similar for negative values and drier than normal conditions. A value of zero indicates an overall normal season, with equally many grid cells falling in the above and below categories. The plot shows the observed average conditions (gray), the forecast from ECMWF (blue) and two ML predictions (red and dark red).}
    \label{fig:expected_tercile_by_year}
\end{figure}

GHA rainfall during the MAM season is also of great interest, but notorious for limited predictability. This is confirmed by the MBSSs shown in Figure~\ref{fig:skill_bootstrapped_chirps}: neither SEAS5 nor our ML models are able to find any source of predictability. The fitted regression coefficients are often zero or very small, the estimated predictive variance accounts for that, and the result is essentially a probabilistic, climatological forecast. This is further confirmed by Figure~\ref{fig:expected_tercile_by_year}, which shows that, especially for the LASSO model with indices-based features, the average regional forecast is quite close to climatology each year.  The performance of the more complex variant with the interaction terms is again on par with the baseline LASSO approach without interaction terms (Figure~\ref{fig:skill_bootstrapped_chirps}). This suggests that for MAM, too, there is no net benefit in the added flexibility, but conversely, the regularization is mostly successful in preventing overfitting a model with many parameters, even in this situation with a very low signal-to-noise ratio. The ECMWF ensemble is here underdispersive, and Figure~\ref{fig:expected_tercile_by_year} shows that the negative skills stems from the forecast having signals that aren't matched by the observations. The data-driven LASSO also has had weak signals for many years, which, similarly, results in a slightly negative overall skill score.   

A similar conclusion can be drawn for the JJAS season, except that here the less complex models have slightly (though not significant) negative skill. A look at the fitted regression coefficients suggests that the ML model more often (compared to MAM) picks up on certain teleconnections but establishes predictor-predictand relations that do not sufficiently generalize out-of-sample to yield positive skill, see also Figure~\ref{fig:expected_tercile_by_year}. For JJAS, the more complex models with interaction terms or EOF-based predictors attain slightly better scores, though at a very low level that is not significantly different from climatology. In this season, the ECMWF forecast is worse than climatology, with the 95th MBSS percentile being (slightly) below zero.

\subsection{More training data vs. more accurate ground truth}\label{subsec:training-sample-size}

\begin{figure}
    \centering
    \includegraphics[width = 0.9\textwidth]{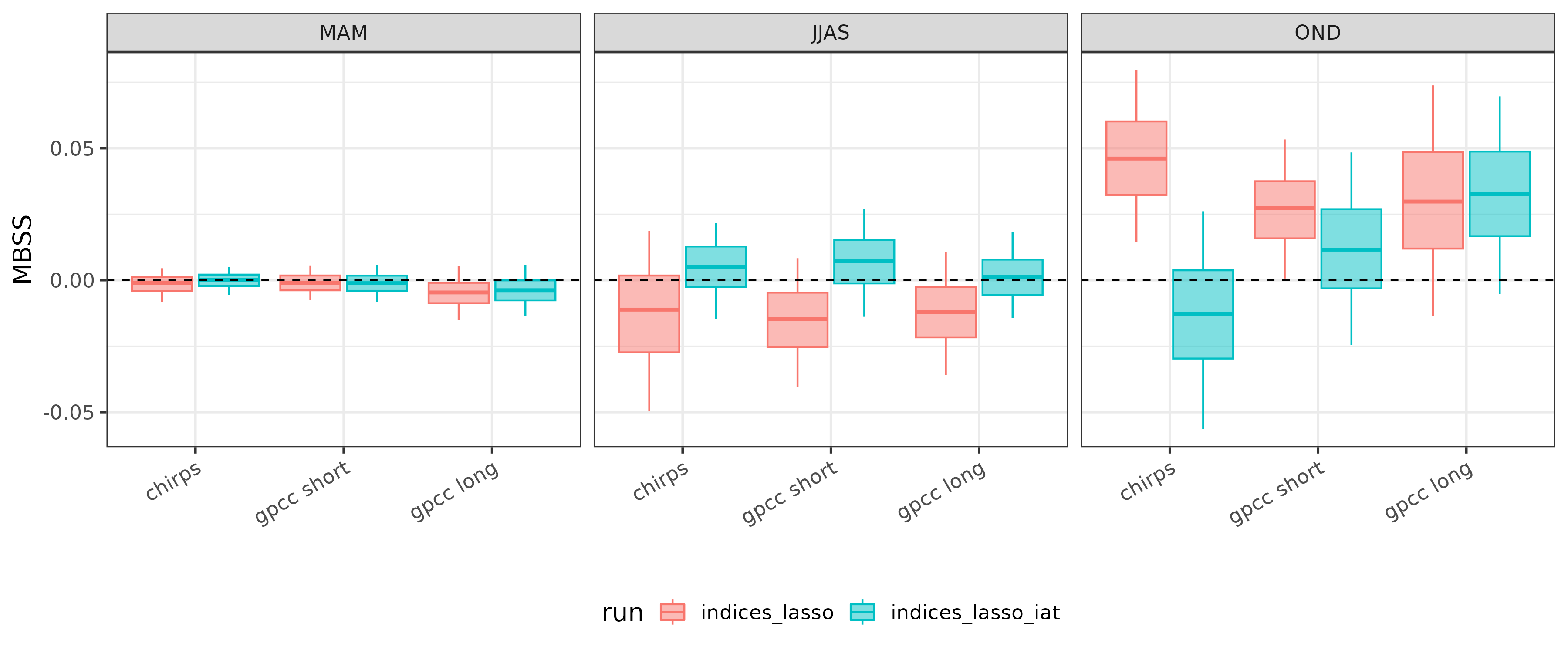}
    \caption{MBSSs for the lasso method with and without interaction terms, using different training datasets.
    The boxes labelled `chirps' are trained on CHIRPS observations 1981-2020, `gpcc short' are trained on the same years using GPCC observations, and `gpcc long' are trained on 1950-2020 using GPCC observations. All forecasts are evaluated over the period 1993-2020 against tercile categories calculated from the same data product that was used for training.}
    \label{fig:skill_bootstrapped_training_lengths}
\end{figure}

We now return to the question of which data set to use as a ground truth and the associated trade-offs between training sample size and data quality. Based on the insights from Sections~\ref{sec:data} and \ref{subsec:ecmwf} we have focused on CHIRPS data. While CHIRPS appears to be a better proxy for observed precipitation than GPCC, it limits the estimation to approximately 40 years of training data. The short training period may, in particular, be the reason for the low skill of the LASSO model with interaction terms in OND, and more data may potentially yield a more stable signal in the MAM and JJAS seasons. To test this, we fit the baseline LASSO model and its variant with interaction terms to GPCC precipitation data over a) the same 1981-2020 training period used above and b) an extended (cross-validated) training period 1950-2020, for which both GPCC and ERA5 are available. As above, the resulting predictions were evaluated over the period 1993-2020, but the evaluation was performed against the respective data product that the ML method was trained against.

\begin{figure}
    \centering
    \includegraphics[width = \textwidth]{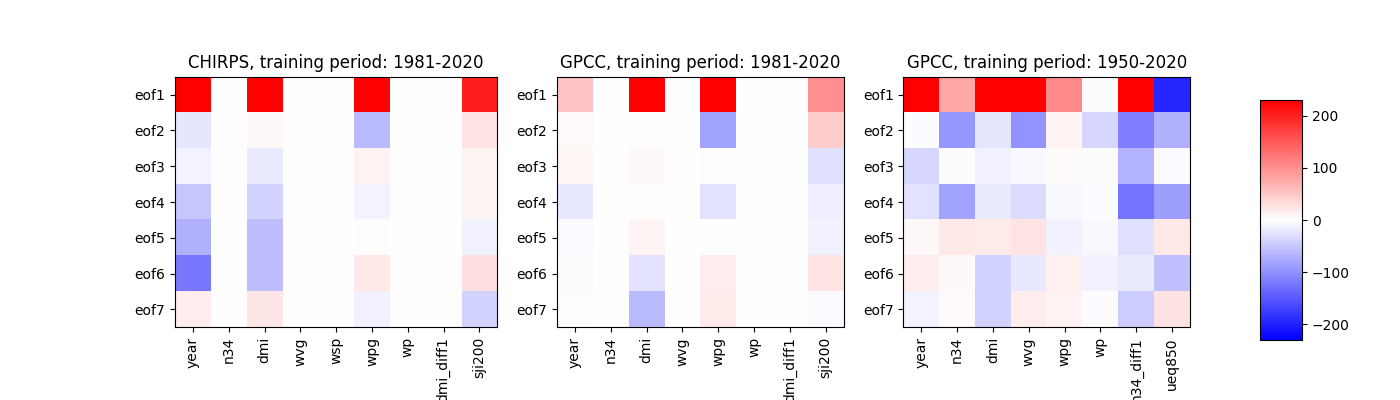}
    \caption{LASSO coefficients fitted for the model predicting OND precipitation amounts based on data available by mid-August. The seven EOFs of the prediction target are on the y-axis, the predictors are on the x-axis with only those included in the plots that passed the respective pre-selection process. Deep red colors are indicative of a strong positive association of a feature with the respective prediction target, deep blue colors indicate a strong negative association.}
    \label{fig:lasso-coefficients-training-comparison}
\end{figure}

Indeed, the relative OND performance of the more complex model with interaction terms improves when more training data is available (see Figure~\ref{fig:skill_bootstrapped_training_lengths}). This improvement, however, is relative to a lower base level of GPCC results compared to CHIRPS results, and the more complex model still only reaches comparable but not better results than the baseline model. The best results are achieved with the model trained and verified against the CHIRPS data, so in our case a more accurate data product is more important than the larger training sample size. Interestingly, the LASSO regression coefficients (see Figure~\ref{fig:lasso-coefficients-training-comparison}) are more similar between the CHIRPS and the GPCC-short experiment than between GPCC-short, and GPCC-long, at least for the most important EOF1. While a temporal trend towards wetter conditions and the role of the SJ200 predictor is more pronounced with the CHIRPS data, both CHIRPS and GPCC-short rely on relatively few predictors, specifically DMI and WPG. When the model is fitted to GPCC data over the long training period, however, additional predictors like WVG and the one-month increase in N34 are considered. The position and size of the corresponding boxes in Figure~\ref{fig:skill_bootstrapped_training_lengths} suggests that the more diverse predictors do not benefit the median performance and even increase the year-to-year variability of the MBSS. On the positive side, it can be noted that the LASSO regularization works as desired and becomes more selective when less training data is available.

For MAM and JJAS, we conclude that the larger training sample size does not help identify a more predictable signal. For MAM, the scores are even slightly worse when the longer training period is used for model fitting. This could be an indication of a non-stationary predictor-predictand relationships, but in all cases the zero skill line is well within the bootstrapped 90\% interval, so the difference may also result from sampling variability.

\subsection{Relation between forecast skill and horizontal resolution}\label{subsec:skill-vs-horizontal-resolution}

\begin{figure}
    \includegraphics[width = 0.94\textwidth, left]{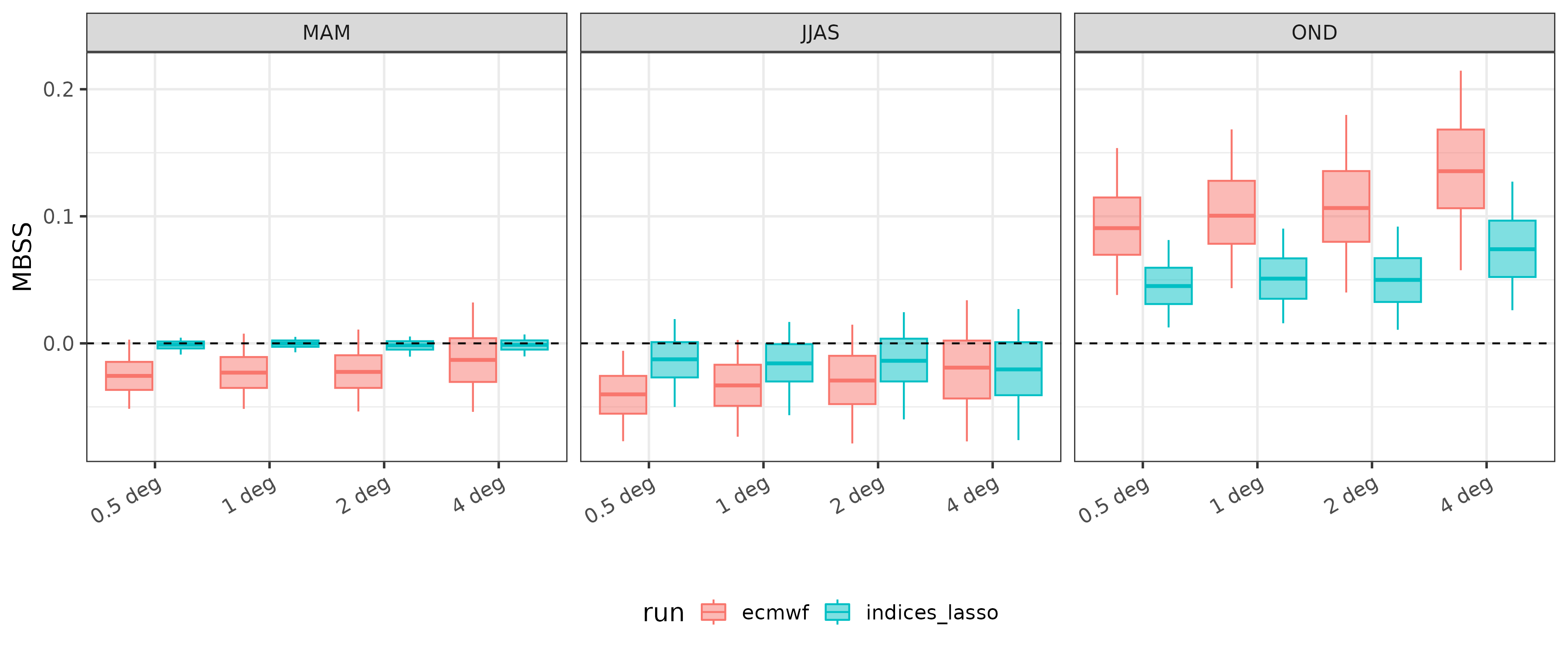}\\
    \includegraphics[width = \textwidth]{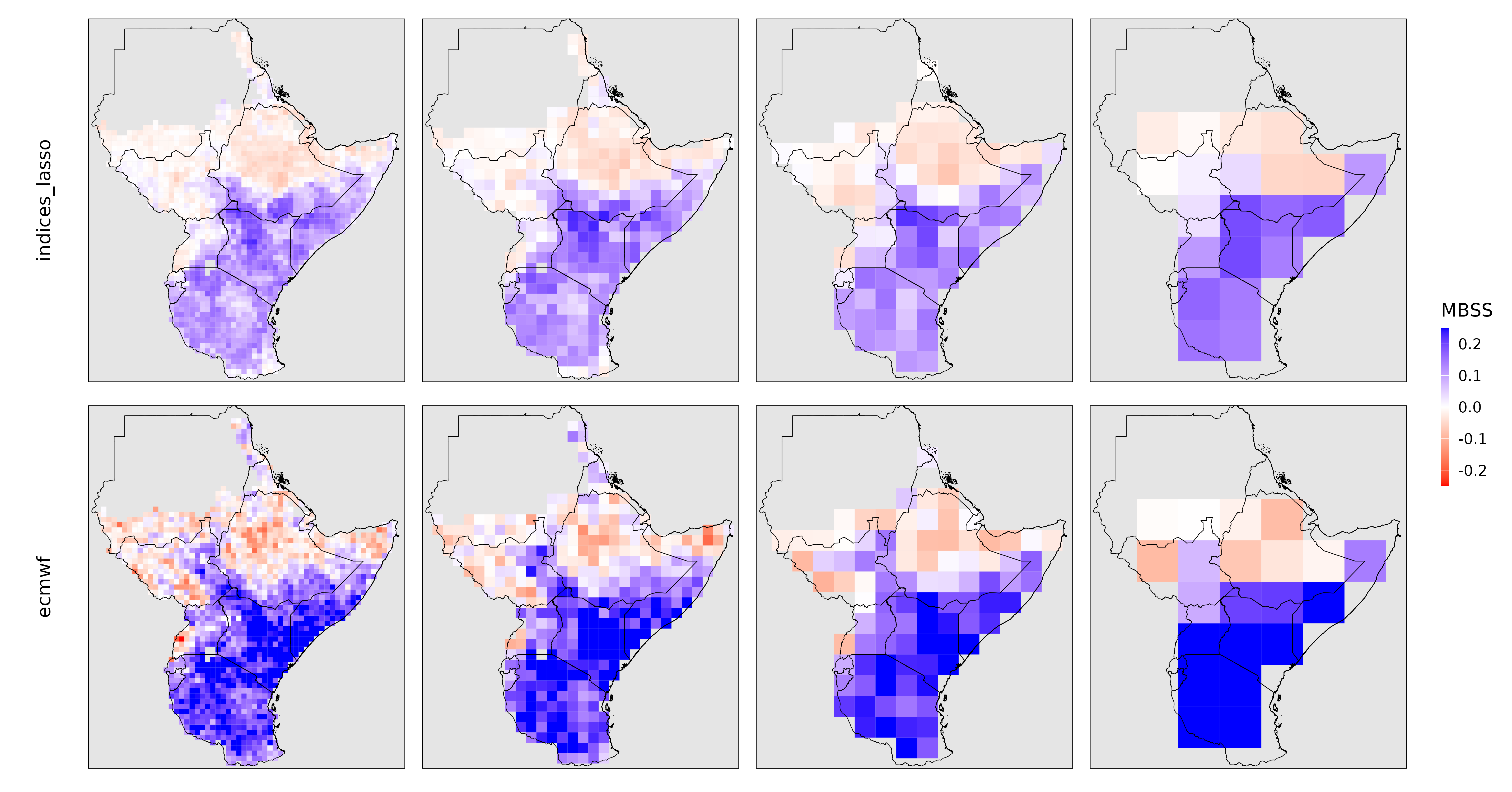}    
    \caption{MBSSs for the LASSO method with indicies as predictors and ECMWF forecasts at different horizontal resolutions. The evaluation is performed against CHIRPS at the respective resolution over the period 1993-2020. 
    }
    \label{fig:skill_bootstrapped_resolutions}
\end{figure}

The results analyzed so far were based on gridded forecasts at a horizontal resolution of 0.5$^\circ\times$0.5$^\circ$. Earlier studies \citep[e.g.][]{Gong&2003} suggest that the skill of seasonal precipitation forecasts improves with increasing spatial aggregation. Here, we investigate this in our setting of tercile forecasts. Specifically, we train and evaluate our baseline ML model (LASSO with indicies as predictors) against CHIRPS data upscaled to a horizontal resolution of 1$^\circ\times$1$^\circ$, 2$^\circ\times$2$^\circ$, and 4$^\circ\times$4$^\circ$, respectively. For this experiment, precipitation amounts are spatially averaged to the coarser grids before calculating the transformed anomalies as described in Section \ref{subsubsec:preprocessing precip}. Coarse gridpoints with partial coverage are included if the CHIRPS data covers at least 50\% of the area contained in the gridpoint. As in all previous experiments, we mask out the driest 25\% of gridpoints, where here the mask is re-calculated for each resolution and season. 

The scores depicted in Figure~\ref{fig:skill_bootstrapped_resolutions} suggest that for MAM and JJAS, the aggregation levels considered here do not make a substantial difference, and that the low predictability of seasonal rainfall amounts remains a challenge at coarser scales. For OND, however, where the results discussed above show some predictability of seasonal rainfall amounts even at the 0.5$^\circ\times$0.5$^\circ$ resolution, the scores obtained with both ECMWF and the ML forecasts further improve with spatial aggregation. This highlights the conflict between the desire of decision-makers for localised information and the scientific realisation that spatial precision comes at the expense of forecast skill, meaning that these two aspects need to be balanced in practice. 

\subsection{Case study}\label{subsec:case-study}

\begin{figure}
    \includegraphics[width=\textwidth]{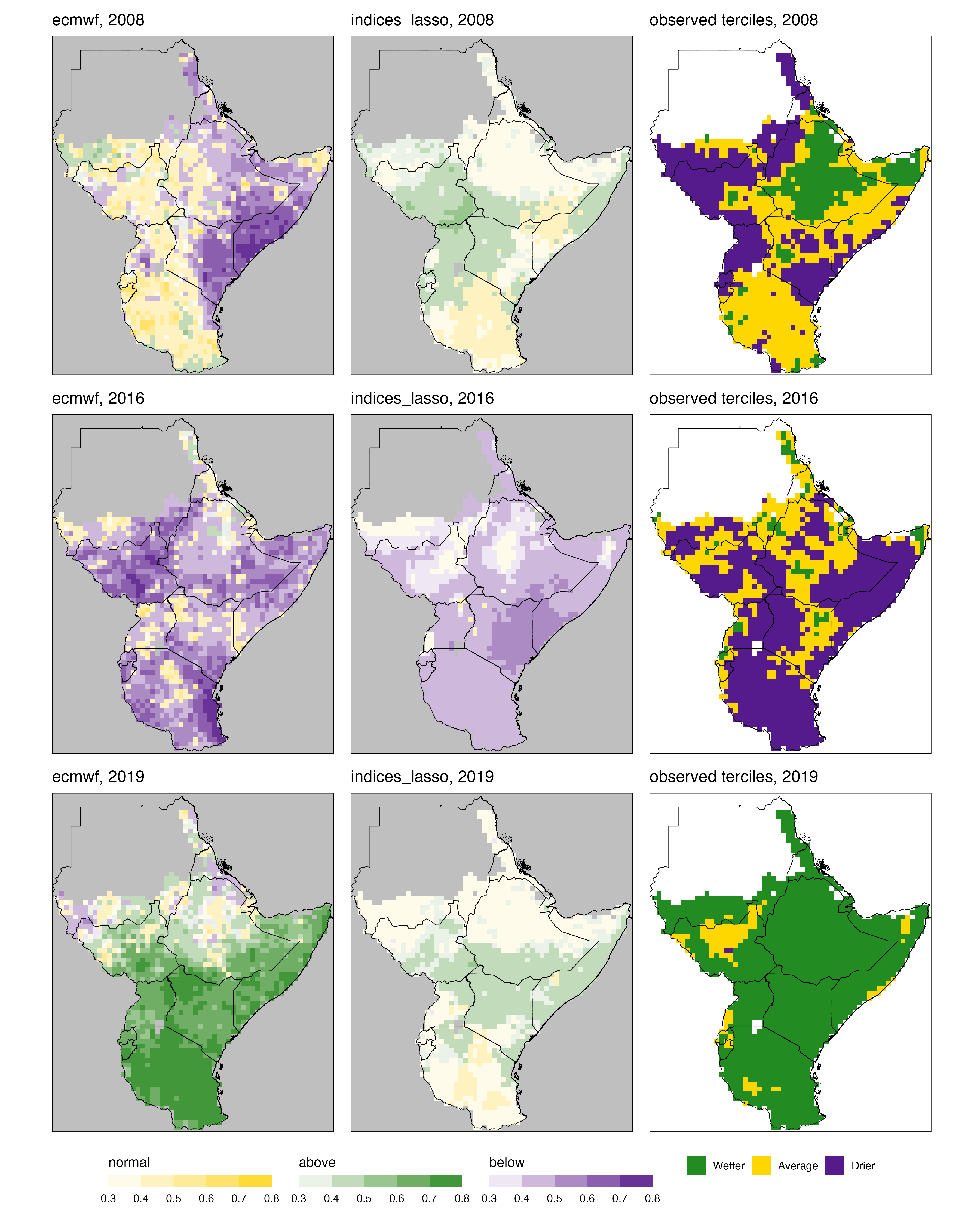}
    \caption{Examples of OND tercile probability forecasts by the ECMWF model and our baseline ML approach, respectively, along with the observed tercile category.}
    \label{fig:probfcst-example}
\end{figure}

To illustrate how specific tercile probability forecast based on the proposed ML look compared to those based on the ECMWF ensemble forecasts and to provide an understanding for when the method works well and when it fails, we depict some of the predictions for selected years, along with the observed category derived from CHIRPS data. These results are shown in Figure~\ref{fig:probfcst-example}. Specifically, we consider the years 2016 and 2019, which were characterized by a negative (2016) and positive (2019) ENSO and IOD during both JJAS and OND, and the year 2008, where ENSO and IOD were in opposite phase in JJAS \citep{Roy&2023}. We saw in Figure~\ref{fig:lasso-coefficients-training-comparison} that DMI, WPG and SJI200 are identified by our ML method as the most important indices for OND rainfall prediction, and we note the the WPG series is strongly correlated with ENSO (not shown here).

In August 2016, DMI was strongly negative, and WPG was negative, which resulted in a clear signal that our ML method could exploit to predict an increased probability of an OND rain deficit over the GHA. This indeed materialized over large parts of the domain. The opposite situation was observed in OND 2019, where almost all parts of GHA received excess rainfall amounts. While the ECMWF model anticipated this situation very well, our ML model predicted the correct tendency but with a considerably weaker signal. A look at the index values in August 2019 reveals that DMI was positive, but WPG was near neutral and SJI200 was negative, thus weakening the `wet signal' of the DMI. In the OND 2008 season, a positive DMI in August led our model to predict an increased probability of wetter than average conditions over large (especially, the western) parts of GHA, but it was mostly a drier than usual OND season that materialized. The teleconnections known to be important drivers of the East African monsoon variability \citep{Roy&2023} thus pointed in a direction that didn't match the outcome, and a model relying on these teleconnection fails to make a useful prediction. Furthermore, we see that even if the ECMWF model correctly predicted the overall tendencies in the region for the OND 2008 season as shown in Figure~\ref{fig:expected_tercile_by_year}, the predicted spatial pattern is quite different from the spatial pattern that materialized. 

\section{Discussion}\label{sec:disc}

Some of the overarching conclusions of the investigations presented in the previous section are matched in similar findings in the literature. \cite{lee2024spring} set up a related modelling framework to predict spring precipitation in South Korea and also find that the more involved elastic net model does not outperform the simpler LASSO. As shown in Figures~\ref{fig:expected_tercile_by_year} and \ref{fig:probfcst-example}, there are several instances of OND forecasts where the ECMWF forecast and the LASSO model predict the same overall tendencies that also materialize, except that the signal is stronger in the ECMWF forecast resulting in this forecast obtaining a better score than the LASSO model. Similarly, in an assessment of subseasonal forecasts for temperature over the contiguous United States, \cite{he2022learning} find that the ML methods they consider yield more conservative magnitudes than a dynamical model. 

In general, however, the comparison of results across studies is complicated by the variability in the setups chosen by the respective authors. As shown in Section~\ref{subsec:case-study}, the results may differ depending on the spatial resolution. Correspondingly, different forecasting targets, such as tercile maps, spatial averages and aggregates, or other indices may vary in underlying predictability even if they refer to the same region and season \citep[e.g.][]{funk2014predicting, Nicholson2014, deman2022seasonal}. For example, \cite{deman2022seasonal} report a positive skill of the SEAS5 ensemble in predicting monthly accumulated precipitation in the Horn of Africa for March, April and May. Furthermore, different evaluation approaches may yield different conclusions regarding overall skill, see e.g. \citet[supplementary material]{mouatadid2023adaptive} regarding the potential difference in skill interpretation for subseasonal dynamical forecasts of temperature and precipitation over the contiguous United States when the skill is measured either in terms of anomaly correlation or the Brier skill score for above normal conditions. More generally, further research is needed to connect standard numercial evaluation measures in weather and climate forecasting with user-relevant information on prediction skill \citep[e.g.][]{landman2023probabilistic}.

The modular design of the machine learning prediction framework shown in Figure~\ref{fig:pipeline} makes it easy to test additional predictors than those discussed above. Inspired by the results of \cite{Maybee&2023}, we investigated, different variants of Madden-Julian Oscillation (MJO) based predictors which summarize the fraction of days in which the MJO is active and in a given phase. Unfortunately, this predictor was not able to improve MAM prediction skill beyond climatological skill, and the results of these experiments are therefore not reported here. %\cite{Vellinga&Milton2018}
The same applies to experiments with tercile probability forecasts for individual months (rather than seasons) as suggested by \citet{camberlin2002east} and \citet{Nicholson2015}, who point out that precipitation amounts during different months of the MAM season are driven by different processes. We did not obtain significant skill for March, April, or May tercile probability forecasts, and have therefore not reported these results in detail.

\section{Conclusions} \label{sec:conclusions} 

Setting up a ML seasonal forecast model as proposed here requires numerous modeling choices, which were systematically studied in order to understand their impact on the forecast quality. Specifically, we conclude the following regarding the research questions stated in the introduction: 
\begin{enumerate}
    \item[(i)-(ii)] Data from more recent years are more accurate due to, e.g., the availability of satellite data. This higher accuracy, along with the potential non-stationarity of teleconnections makes a case for preferring CHIRPS over GPCC data as ground truth, even if this halves the available training data and constrains the complexity of the ML model. However, we find that all available CHIRPS data should be used. Alternatively, for more data-intensive ML algorithms, the training data may potentially be augmented by training on dynamical model output \citep{gibson2021training}. 
    \item[(iii)] We investigate including predictor interactions that would permit, for example, a different relation of DMI with GHA rainfall depending on whether ENSO is positive or negative. We find that the detrimental effects of the associated increase in model parameters outweigh the benefits of added flexibility. The proposed inclusion of interaction terms would have retained interpretability, but is generally unable to improve model performance. Similarly, using elastic net regularization rather than LASSO does not notably improve model performance. 
    \item[(iv)] The skill of indices-based predictions are found to be comparable to using data-driven predictor constructions. In particular, we are not able to identify novel sources of predictability based on the data-driven approach.
    \item[(v)] Our results confirm earlier conclusions \citep[e.g.][]{Gong&2003} that the skill of seasonal precipitation forecasts improves with increasing spatial aggregation. 
\end{enumerate}

\section*{Acknowledgements}
This work had benefited greatly from insightful discussions with Edward Pope, Andrew Coleman and Elisabeth Thompson at the UK MET Office, Erik Kolstad and Stefan Sobolowski at NORCE. This research was supported by the European Union's Horizon 2020 research and innovation programme under grant agreement no. 869730 (CONFER). All data used in this study is publicly available: ERA5 reanalysis data and SEAS5 ensemble forecasts can be downloaded from the Copernicus Climate Change Service Climate Data Store (\url{https://cds.climate.copernicus.eu/}), CHIRPS data is made available at \url{https://data.chc.ucsb.edu/products/CHIRPS-2.0/}, and GPCC data can be downloaded at \url{https://opendata.dwd.de/climate_environment/GPCC/full_data_monthly_v2022/}. A version of the machine learning methodology proposed in this paper that is more oriented towards operational use (e.g., no cross-validation of training data) is included in a Python package accessible at \url{https://github.com/SeasonalForecastingEngine/CONFER-WP3}.

%\bibliographystyle{apalike}
%\bibliography{refs.bib}

\appendix

\section{Indices}
\begin{longtable}{|c|c|p{3cm}|p{6cm}|c|}
  \caption{Atmospheric and SST-based indices used as potential predictors for rainfall over GHA\label{tab:indices}} \\
  \hline
Handle & Index & Reference & Definition/Location & Level (mb) \\
\hline
\endfirsthead
\multicolumn{5}{c}%
{{\bfseries \tablename\ \thetable{} -- continued from previous page}} \\
\hline
Handle & Index & Reference & Definition/Location & Level (mb) \\
\hline
\endhead
\hline
\multicolumn{5}{|r|}{{Continued on next page}} \\
\hline
\endfoot
\hline
\endlastfoot

n34 & El Ni\~{n}o 3.4 & {\citet{trenberth1997definition}} & Standardized SST anomalies averaged over \(5^\circ \text{S}-5^\circ \text{N}\); \(170^\circ \text{W}-120^\circ \text{W}\) & SST \\
\hline
n3 & El Ni\~{n}o 3 & & Standardized SST anomalies averaged in a region bounded by \(5^\circ \text{S}-5^\circ \text{N}\), \(150^\circ \text{W}-90^\circ \text{W}\) & SST \\
\hline
n4 & El Ni\~{n}o 4 & & Standardized SST anomalies averaged in region bounded by \(5^\circ \text{N}-5^\circ \text{S}\), \(160^\circ \text{E}-150^\circ \text{W}\) & SST \\
\hline
wpg & WPG & {\citet{funk2014predicting}} & Western Pacific Gradient, difference between standardized SST anomalies in the El Ni\~{n}o 4 region and the western Pacific (WP; \(0^\circ \text{N}-20^\circ \text{N}\), \(130^\circ \text{E}-150^\circ \text{E}\)) & SST \\
\hline
dmi & DMI & {\citet{saji1999dipole}} & Dipole Mode Index as the difference between standardized SST anomalies over the western IO (WIO; \(10^\circ \text{S}-10^\circ \text{N}\), \(50^\circ \text{E}-70^\circ \text{E}\)) and eastern IO (EIO; \(10^\circ \text{S}-0^\circ \text{S}\), \(90^\circ \text{E}-110^\circ \text{E}\)) & SST \\
\hline
ueq850 & UEQ850 & {\citet{hastenrath2011circulation}} & Standardized zonal wind averaged over the central equatorial Indian Ocean \((4^\circ \text{N}-4^\circ \text{S}, 60^\circ \text{E}-90^\circ \text{E})\)  & 850\\
\hline
ueq200 & UEQ200 & {\citet{hastenrath2011circulation}} & \((4^\circ \text{N}-4^\circ \text{S}, 60^\circ \text{E}-90^\circ \text{E})\) is an index of the zonal component of surface wind over the central equatorial Indian Ocean & 200 \\
\hline
wp & WP & {\citet{Funk&2018}} & Standardized SST anomalies averaged in region bounded by \(120^\circ \text{E}-160^\circ \text{E}\), \(15^\circ \text{S}-20^\circ \text{N}\) & SST \\
\hline
wnp & WNP & {\citet{Funk&2018}} & Standardized SST anomalies averaged in region bounded by \(160^\circ \text{E}-150^\circ \text{W}\), \(20^\circ \text{N}-35^\circ \text{N}\) & SST \\
\hline
wsp & WSP & {\citet{Funk&2018}} & Standardized SST anomalies averaged in region bounded by \(155^\circ \text{E}-150^\circ \text{W}\), \(15^\circ \text{S}-30^\circ \text{S}\) & SST \\
\hline
wvg & WVG & {\citet{Funk&2018}} & Western V-Gradient, calculated as \(n4 -(wp + wnp + wsp)/3\) & SST \\
\hline

\end{longtable}

\end{document}